\documentclass[reprint, superscriptaddress, amsmath,amssymb, aps, prb]{revtex4-2}
\usepackage{graphicx}
\usepackage{dcolumn}
\usepackage{bm}
\usepackage{hyperref}

\begin{document}

\title{RKKY interaction in helical higher-order topological insulators}

\author{Sha Jin}
\affiliation{School of Science, Chongqing University of Posts and Telecommunications, Chongqing 400065, China}
\author{Jian Li}
\affiliation{School of Science, Chongqing University of Posts and Telecommunications, Chongqing 400065, China}
\affiliation{Institute for Advanced Sciences, Chongqing University of Posts and Telecommunications, Chongqing 400065, China}
\affiliation{Southwest Center for Theoretical Physics, Chongqing University, Chongqing 401331, China}
\author{Qing-Xu Li}
\affiliation{School of Science, Chongqing University of Posts and Telecommunications, Chongqing 400065, China}
\affiliation{Institute for Advanced Sciences, Chongqing University of Posts and Telecommunications, Chongqing 400065, China}
\affiliation{Southwest Center for Theoretical Physics, Chongqing University, Chongqing 401331, China}
\author{Jia-Ji Zhu}
\email{zhujj@cqupt.edu.cn}
\affiliation{School of Science, Chongqing University of Posts and Telecommunications, Chongqing 400065, China}
\affiliation{Institute for Advanced Sciences, Chongqing University of Posts and Telecommunications, Chongqing 400065, China}
\affiliation{Southwest Center for Theoretical Physics, Chongqing University, Chongqing 401331, China}

\date{\today}

\begin{abstract}
We theoretically investigate the RKKY interaction in helical higher-order topological insulators (HOTIs), revealing distinct behaviors mediated by hinge and Dirac-type bulk carriers. Our findings show that hinge-mediated interactions consist of Heisenberg, Ising, and Dzyaloshinskii-Moriya (DM) terms, exhibiting a decay with impurity spacing $z$ and oscillations with Fermi energy $\varepsilon_F$. These interactions demonstrate ferromagnetic behaviors for the Heisenberg and Ising terms and alternating behavior for the DM term. In contrast, bulk-mediated interactions include Heisenberg, twisted Ising, and DM terms, with a conventional cubic oscillating decay. This study highlights the nuanced interplay between hinge and bulk RKKY interactions in HOTIs, offering insights into the design of next-generation quantum devices based on the HOTIs.
\end{abstract}


\maketitle

\section{Introduction}
The higher-order topological insulators (HOTIs) describe topological materials of $d$-dimensional insulated bulk and ($d-n$)-dimensional gapless boundary states\cite{kim2020recent, xie2021higher}. There are zero-dimensional gapless corner states in two-dimensional (2D) second-order topological insulators\cite{benalcazar2017quantized,liu2017novel,ZhiKangLin74302,HuanChen17202} and three-dimensional (3D) third-order topological insulators\cite{ni2020demonstration,liu2020octupole}, and one-dimensional (1D) gapless hinge states in 3D second-order topological insulators\cite{hackenbroich2021fractional,wang2021structural,fu2021bulk}. The HOTI can be understood as a special topological crystalline insulator (TCI). For example, one can turn SnTe, a well-known topological crystalline insulator, into a HOTI by reducing temperature or applying uniaxial strain to gap out the Dirac cones of surfaces\cite{schindler2018higher}. Compared with the TCI, the HOTI possesses higher symmetry beyond the crystalline symmetry. For instance, a pristine TCI, the cubic SnTe, is protected by mirror symmetry\cite{hsieh2012topological,tanaka2012experimental}. In contrast, the related HOTI, the strained octahedral SnTe, is protected by both mirror symmetry and time-reversal symmetry\cite{schindler2018higher}. The topological invariants of the HOTIs can be interpreted physically by the electric multipole moments, which further extends the dipole moment theory in TCIs\cite{benalcazar2017electric}. Like the ordinary topological insulator (TI) with robust ($d-1$)-dimensional boundary states, the HOTIs possess robust ($d-n$)-dimensional boundary states which can also survive defects, impurities, or other perturbations. For instance, the Pb$_{0.67}$Sn$_{0.33}$Se bulk crystal holds 1D nontrivial hinge states with a striking robustness to defects, strong magnetic fields, and elevated temperatures\cite{Sessi2016RobustSM}. The robustness to perturbations makes HOTIs a promising material for designing high-stable electronic or spintronic devices.

One can break or preserve the time-reversal symmetry to switch the 1D hinge state in 3D second-order topological insulators between chiral and helical regimes. The chiral hinge states are electrons propagating unidirectionally like the edge states of the 2D quantum Hall effect or the quantum anomalous Hall effect. The helical hinge states are Kramers pairs counter-propagating like the edge states of a 2D quantum spin Hall effect\cite{bernevig2006quantum}, which are also spin-momentum locked and free from back-scattering. The unique spin-momentum locking and counter-propagating currents of helical hinge states allow the design of a helical nanorod which has exactly $n$-channels of ballistic transport\cite{fang2019new} and the spin manipulation based on spin-momentum locking\cite{kohda2019spin,yang2020spin}. The helical hinge modes may have different configurations; for example, there are two possible helical hinge mode configurations in Bi\cite{aggarwal2021evidence}, where one configuration has $C_2$ symmetry and time-reversal symmetry and the other only has the time-reversal symmetry. The various configurations of helical states offer the possibility of designing devices by different helical circuits. The helical hinge states have already been realized in materials such as bismuth\cite{schindler2018bismuth}, SnTe\cite{schindler2018higher}, $\alpha$-Bi$_{4}$Br$_{4}$\cite{hsu2019purely,shumiya2022evidence} and MoTe$_{2}$\cite{wang2019higher}. The helical modes can also be achieved by methods of artificial manipulation, e.g., using bismuth-halide chains by the van der Waals stacking\cite{noguchi2021evidence}, an array of weakly tunnel-coupled Rashba nanowires\cite{laubscher2023fractional}, or a $C_{6}$-symmetric topological crystalline meta-material based on the acoustic samples\cite{yang2020helical}. 

Ruderman-Kittel-Kasuya-Yosida (RKKY) interaction describes the exchange coupling between magnetic impurities mediated by the itinerant carriers in the host material. The RKKY interaction has been extensively investigated in various systems such as low-dimensional quantum structures\cite{craig2004tunable,usaj2005tuning}, graphene and Dirac semimetals\cite{dugaev2006exchange,brey2007diluted}, topological insulators\cite{liu2009magnetic,zhu2011electrically} and Weyl semimetals\cite{chang2015rkky,hosseini2015ruderman}, and topological crystalline insulators\cite{yarmohammadi2020effective,cheraghchi2021anisotropic,yarmohammadi2023noncollinear}. The boundary effects in topological materials are predicted to be highly intriguing. HOTIs possess 1D topological hinge states that may offer a unique RKKY interaction resilient to perturbations. Additionally, varying configurations of helical hinges in HOTIs can serve as a versatile platform for magnetic switching through RKKY interaction.

In this work, we focus on the RKKY interaction between magnetic impurities positioned in the helical hinge of a 3D second-order TI. Here, we carefully analyze the system and describe the RKKY interaction not only mediated by the hinge states but also by the bulk states. By utilizing Green's function technique and the low-energy effective Hamiltonian of a 3D second-order TI\cite{schindler2018higher}, we can arrive at the analytical expressions of the hinge and bulk RKKY interactions. The hinge RKKY interaction consists of three terms: the conventional Heisenberg type interaction, the Dzyaloshinskii-Moriya (DM) type interaction along the certain direction, and the Ising type interaction perpendicular to the hinge. The strength of the hinge RKKY interaction is linearly proportional to the reciprocal of impurity spacing and shows a sinusoidal pattern of the product of Fermi energy and impurity spacing. The hinge RKKY interaction exhibits two distinct branches, with the only difference being the sign of the DM interaction at equal impurity spacing. The opposite sign of the DM term reflects the helical nature of the hinge states. The bulk RKKY interaction consists of the conventional Heisenberg-type, twisted Ising-type and DM-type interactions. The strength of the bulk RKKY interaction decreases with the impurity spacing $R^3$ and also shows a sinusoidal pattern related to Fermi energy and impurity spacing. The interplay between the hinge and bulk RKKY interaction allows for developing several quantum devices with varying hinge setups.

\section{Model}
\begin{figure}[thp]
\begin{center}
\includegraphics[width=\columnwidth]{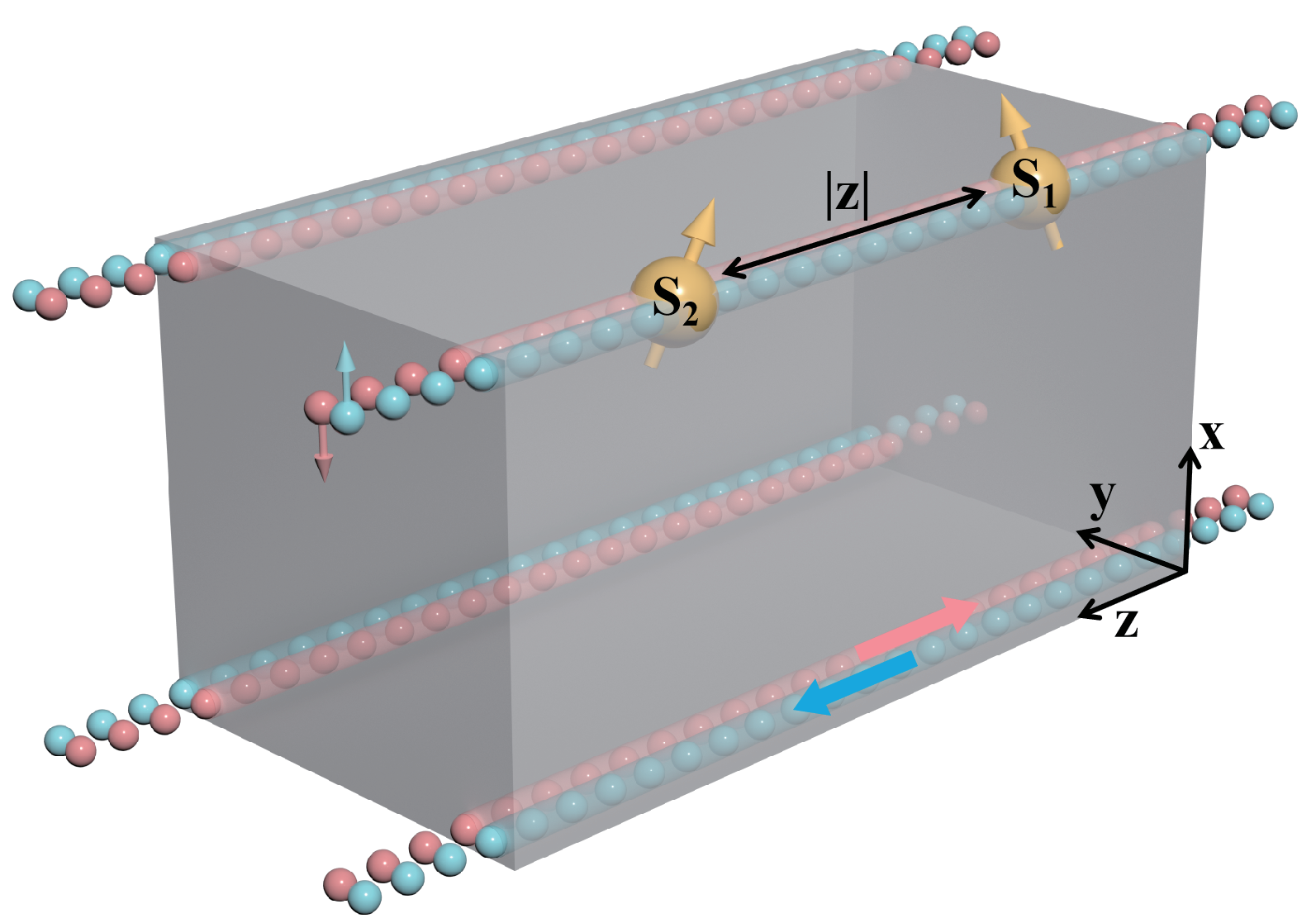}\\
\parbox[c]{\columnwidth}{\caption{
The diagram of magnetic impurities in the helical hinge of HOTIs. The periodic boundary conditions in the $z$ direction are assumed. The insulated bulk and surface states are painted as gray, while the helical hinge states consist of Kramers pairs are denoted by colorful balls. The blue (pink) balls stand for carriers with spin up (down). Blue and pink arrows along the $z$ axis mark the direction of motion of spin-up and spin-down carriers. The Large yellow balls denote magnetic impurities with magnetic orientation.}\label{fig:diagram}}
\end{center}
\end{figure}

We consider a helical HOTI, which has a square cross-section in $xy$ plane but periodic boundary conditions in $z$ direction as shown in Fig.~\ref{fig:diagram}. The bulk and surface states are insulated, while the helical hinge states along the $z$ direction are conductive of the counter-propagating helical Kramers pairs. The magnetic impurities are located at the hinges. 

The Hamiltonian of itinerant carriers from the helical hinge state is\cite{schindler2018higher}
\begin{eqnarray}
\mathcal{H}(k_{z})&=&
\rho_{0}\tau _{x}\sigma _{x}(-{\rm i}\partial _{x})+\rho_{0}\tau _{x}\sigma _{y}(-{\rm i}\partial _{y})+\rho_{0}\tau _{x}\sigma_{z}k_{z}\notag\\
&&+\rho_{0}\tau _{z}\sigma _{0}\delta_{1}+\rho_{z}\tau_{y}\sigma _{0}\delta_{2},
\label{eq:helicalH}
\end{eqnarray}
where $\sigma _{0}$ is the unit matrix and $\sigma_{i}$ ($i=x,y,z$) are the Pauli matrices acting on spin degree of freedom, $\tau_{i}$ the Pauli matrices acting on the $d$ and $f$ orbits, and $\rho_{0}$ the unit matrix and $\rho_{i}$ the Pauli matrices acting on the $d_{x^{2}-y^{2}}(f_{(x^{2}-y^{2})z})$ and $ d_{xy}(f_{xyz})$ orbitals respectively. The basis of the Hamiltonian (\ref{eq:helicalH}) is $(|d_{x^{2}-y^{2}},\uparrow\rangle,|d_{x^{2}-y^{2}},\downarrow\rangle,|f_{(x^{2}-y^{2})z},\uparrow\rangle,|f_{(x^{2}-y^{2})z},\downarrow\rangle,|d_{xy},\uparrow\rangle,|d_{xy},\downarrow\rangle,|f_{xyz},\uparrow\rangle,|f_{xyz},\downarrow\rangle)^{T}$. Here $ \delta_{1}=(x+y)/\sqrt{2}$ and $ \delta_{2}=(-x+y)/\sqrt{2}$ form a vortex with winding number $1$\cite{schindler2018higher}. Then the Eq.~(\ref{eq:helicalH}) can be written in a matrix form: 
\begin{eqnarray}
\mathcal{H}(k_{z})=
\left( 
\begin{array}{cc}
\mathcal{H}_{1}(k_{z}) & 0 \\ 
0 & \mathcal{H}_{2}(k_{z})
\end{array}
\right) 
\end{eqnarray}
where
\begin{eqnarray}
\mathcal{H}_{1}(k_{z})=
\left( 
\begin{array}{cccc}
\delta _{1} & 0 & k_{z}-{\rm i}\delta _{2} & -{\rm i}\partial _{x}-\partial _{y} \\ 
0 & \delta _{1} & -{\rm i}\partial _{x}+\partial _{y} & -k_{z}-{\rm i}\delta _{2} \\ 
k_{z}+{\rm i}\delta _{2} & -{\rm i}\partial _{x}-\partial _{y} & -\delta _{1} & 0 \\ 
-{\rm i}\partial _{x}+\partial _{y} & -k_{z}+{\rm i}\delta _{2} & 0 & -\delta _{1} \\ 
\end{array}
\right)
\end{eqnarray}
and
\begin{eqnarray}
\mathcal{H}_{2}(k_{z})=
\left( 
\begin{array}{cccc}
\delta _{1} & 0 & k_{z}+{\rm i}\delta _{2} & -{\rm i}\partial_{x}-\partial _{y} \\ 
0 & \delta _{1} & -{\rm i}\partial _{x}+\partial _{y} & -k_{z}+{\rm i}\delta _{2} \\ 
k_{z}-{\rm i}\delta _{2} & -{\rm i}\partial _{x}-\partial _{y} & -\delta_{1} & 0 \\ 
-{\rm i}\partial _{x}+\partial _{y} & -k_{z}-{\rm i}\delta _{2} & 0 & -\delta _{1}
\end{array}
\right).
\end{eqnarray}

We apply a similarity transformation $ R=\mathrm{exp}(\frac{{\rm i}\pi}{4}\rho_{z}\tau_{y}\sigma _{0}) $ on the Hamiltonian (\ref{eq:helicalH}), and chose $ k_{z}=0 $. After an elementary transformation, the matrix form becomes a block anti-diagonal matrix form
\begin{eqnarray}
H(k_{z}=0)=\left(
\begin{array}{cccc}
0 & 0 & 0 & h_{1} \\
0 & 0 & h_{2} & 0 \\
0 & h_{2}^{\dagger } & 0 & 0 \\
h_{1}^{\dagger } & 0 & 0 & 0
\end{array}
\right).
\label{eq:off_diagonal_H}
\end{eqnarray}
with
\begin{eqnarray*}
h_{1}&=
\left( 
\begin{array}{cc}
 {\rm i}\partial _{x}+\partial _{y} & \delta _{1}+{\rm i}\delta_{2}\\ 
\delta _{1}-{\rm i}\delta _{2} & -{\rm i}\partial _{x}+\partial_{y}
\end{array}
\right),\\
h_{2}&=
\left( 
\begin{array}{cc}
{\rm i}\partial _{x}-\partial _{y} & -\delta _{1}+{\rm i}\delta _{2}\\ 
-\delta _{1}-{\rm i}\delta _{2} & -{\rm i}\partial _{x}-\partial _{y}
\end{array}
\right),\\
h_{2}^{\dagger }&=
\left( 
\begin{array}{cc}
{\rm i}\partial _{x}+\partial _{y} & -\delta _{1}+{\rm i}\delta _{2}\\ 
-\delta _{1}-{\rm i}\delta _{2} & -{\rm i}\partial _{x}+\partial _{y}
\end{array}
\right),\\
h_{1}^{\dagger }&=
\left( 
\begin{array}{cc}
{\rm i}\partial _{x}-\partial _{y} & \delta _{1}+{\rm i}\delta _{2} \\ 
\delta _{1}-{\rm i}\delta _{2} & -{\rm i}\partial _{x}-\partial _{y}
\end{array}
\right).
\end{eqnarray*}
Here we should remind that $ (\partial _{x})^{\dagger }=-\partial _{x} $, $ (\partial _{y})^{\dagger }=-\partial _{y} $, $ \delta_{1}=(x+y)/\sqrt{2} $ and $ \delta_{2}=(-x+y)/\sqrt{2} $ .

The Hamiltonian $H(k_{z})$ has Kramers degenerate zero modes propagating along the hinge when $ k_{z}=0 $\cite{schindler2018higher}. We can solve the eigenequation $ H(k_{z}=0)(\phi_{1},\phi_{2},\phi_{3},\phi_{4},\phi_{5},\phi_{6},\phi_{7},\phi_{8})^{T}=0$ and decompose it into a series of equations
\begin{eqnarray}
\left\{ 
\begin{array}{c}
({\rm i}\partial _{x}-\partial _{y})\phi _{1}+\frac{1-{\rm i}}{\sqrt{2}}(x+{\rm i}y)\phi _{2}=0 \\ 
\frac{1+{\rm i}}{\sqrt{2}}(x-{\rm i}y)\phi _{1}-({\rm i}\partial _{x}+\partial _{y})\phi _{2}=0
\end{array}
\right. .
\end{eqnarray}
Then we can get the separate equation for $\phi_1$
\begin{eqnarray}\label{eq:partial_derivative}
\left( x-{\rm i}y\right) \phi _{1}-\left( \partial _{x}-{\rm i}\partial _{y}\right) 
\left[ \frac{1}{x+{\rm i}y}\left( \partial _{x}+{\rm i}\partial _{y}\right) \phi _{1}\right] =0,
\end{eqnarray}
and arrive at the equation
\begin{eqnarray}
&x\partial^2 _{x}X\left( x\right) Y\left( y\right)+x\partial^2_{y}X\left( x\right) Y\left( y\right)- 2\partial _{x}X\left(x\right) Y\left( y\right)\notag\\
&-x\left( x^{2}+y^{2}\right) X\left( x\right)Y\left( y\right)=0
\end{eqnarray}
by setting $ \phi _{1}=X\left( x\right) Y\left( y\right) $.

After some standard derivations of solving the equation, we can get the solutions as  
\begin{eqnarray}
X\left( x\right)=Y\left( x\right)=\frac{\left(c_{1}-{\rm i}c_{2}\right)}{2} e^{-\frac{x^{2}}{2}}+\frac{\left(c_{1}+{\rm i}c_{2}\right)}{2} e^{\frac{x^{2}}{2}}
\end{eqnarray}
with constants $ c_1$ and $ c_2$. We have to discard the $ \exp(x^2/2)$ part for the sake of the physical reason of the wave function. Then the $\phi _{1}\left( x,y\right) =X(x)Y(y)=ce^{-\frac{x^{2}+y^{2}}{2}}$ with complex constant $ c $, and similarly $\phi _{2}=c\frac{\left({\rm i}-1\right) }{\sqrt{2}}e^{-\frac{x^{2}+y^{2}}{2}}$.  Furthermore, the eigenmodes of $H(k_{z}=0)$ can be derived by computing the remaining equations. These will consist of two counter-propagating Kramers paired eigenmodes: 
\begin{eqnarray}\label{eq:eigenstates}
\psi _{1}=ce^{-\frac{x^{2}+y^{2}}{2}}\left( 1,\frac{\left({\rm i}-1\right) }{%
\sqrt{2}},0,0,0,0,0,0\right)^{T}, \\
\psi _{2}=ce^{-\frac{x^{2}+y^{2}}{2}}\left( 0,0,\frac{\left({\rm i}-1\right) }{%
\sqrt{2}},1,0,0,0,0\right)^{T}
\end{eqnarray}
with $c$ being the appropriate normalization factor. The dispersion about $k_{z}$ can be inferred from the matrix elements 
\begin{eqnarray}
\left( 
\begin{array}{cc}
\left\langle\psi _{1}\right\vert H\left( k_{z}\right) \left\vert \psi
_{1}\right\rangle  & \left\langle \psi _{1}\right\vert H\left( k_{z}\right)
\left\vert \psi _{2}\right\rangle \\ 
\left\langle\psi _{2}\right\vert H\left( k_{z}\right) \left\vert \psi
_{1}\right\rangle  & \left\langle \psi _{2}\right\vert H\left( k_{z}\right)
\left\vert \psi _{2}\right\rangle
\end{array}
\right) 
=
\left( 
\begin{array}{cc}
-k_{z} & 0 \\ 0 & k_{z}
\end{array}
\right). 
\end{eqnarray}
The eigenmode $\left\vert\psi_{+}\right\rangle =\left\vert\psi_{2}\right\rangle $ when eigenvalue $\epsilon =k_{z}$ and $\left\vert\psi_{-}\right\rangle =\left\vert\psi_{1}\right\rangle $ when $\epsilon =-k_{z}$. 

In the presence of magnetic impurities within the helical hinge, the interaction between impurity $\bm{S}_{i}$ and itinerant carriers $\bm{\sigma}$ can be expressed as
\begin{eqnarray}\label{eq:intH}
H_{i}^{\mathrm{int}}=
J (\tau_0+\tau_x) \otimes ( \bm{\sigma} \cdot \bm{S}_{i})\delta(z-z_{i}),
\end{eqnarray}
where constant $J$ is coupling strength and $\tau_0 $ and $\tau_x $ are matrices acting on the same orbitals as in Eq.~(\ref{eq:helicalH}). Here, $\tau_0+\tau_x$ means that both the $d$ and $f$ orbitals contribute to the interaction.  $\bm{\sigma}=(\sigma_{x},\sigma_{y},\sigma_{z})$ is for the spin vector of carriers and $\bm{S}_{i}=(S_{ix},S_{iy},S_{iz})$ for the spin vector of impurities. $\delta(z-z_{i})$ is Dirac $\delta$-function which means such interaction is a short-range contact exchange interactions\cite{hickel2004proper,nolting2001low}. 

\section{RKKY interaction in helical hinges} 
First, we consider the RKKY interaction mediating by itinerant carriers between two impurities with localized spin in the hinge of helical HOTIs. The RKKY interaction between impurity 1 and 2 can be calculated by 
\begin{eqnarray}\label{eq:RKKY}
H_{1,2}^{\mathrm{RKKY}}=-\frac{1}{\pi } \mathrm{Im}\int_{-\infty }^{\varepsilon_{F}}\mathrm{Tr}\left[H_{1}^{int}G_{1}(z,\varepsilon) H_{2}^{int}G_{2}(-z,\varepsilon)\right] d\varepsilon,
\end{eqnarray}
where $\varepsilon_{F}$ is Fermi energy and $G(z,\varepsilon)$ is the real space Green's function of the itinerant carriers. 

Note that the momentum $k_x$ and $k_y$ are not good quantum numbers, and we cannot write the momentum space Green's function directly. However, we can write the Green's function in the eigenspace of $H(k_{z})$ instead. For simplicity, we rewrite the eigenstates Eq.~(\ref{eq:eigenstates}) in the subspace ${\bm \tau} \otimes {\bm \sigma}$, and the corresponding basis are $(\tau _{+}\uparrow ,\tau _{+}\downarrow ,\tau _{-}\uparrow ,\tau _{-}\downarrow )^{T}$, where $\tau _{+(-)}$ stands for $d(f)$ orbitals and $ \uparrow (\downarrow)$ stands for spin up (down). The rewritten eigenstates are
\begin{eqnarray}
\left\vert\psi_{+}\right\rangle =& Ne^{-\frac{x^{2}+y^{2}}{2}} \left\vert \tau _{-} \right\rangle \left( \frac{\left({\rm i}-1\right) }{\sqrt{2}}\left\vert \uparrow \right\rangle +\left\vert\downarrow \right\rangle \right), \\
\left\vert\psi_{-}\right\rangle =& Ne^{-\frac{x^{2}+y^{2}}{2}} \left\vert \tau _{+} \right\rangle \left( \left\vert\uparrow \right\rangle +\frac{\left( {\rm i}-1\right) }{\sqrt{2}}\left\vert\downarrow \right\rangle \right).
\end{eqnarray}
In the eigenspace of $H(k_{z})$ the Green's function is
\begin{eqnarray}
G(k_{z},\varepsilon )=
\frac{\psi _{+}\left( x,y\right) \psi _{+}^{\dag }\left( x^{\prime},y^{\prime }\right) }{(\varepsilon+{\rm i}0^{+})-k_{z}}
+\frac{\psi _{-}\left(x,y\right) \psi _{-}^{\dag }\left( x^{\prime},y^{\prime }\right) }{(\varepsilon+{\rm i}0^{+})+k_{z}}. 
\end{eqnarray}
The corresponding matrix form is
\begin{eqnarray}
G(k_{z},\varepsilon )&=&
\left( 
\begin{array}{cccc}
G_{\tau _{+}\uparrow \tau _{+}\uparrow } & G_{\tau _{+}\uparrow \tau _{+}\downarrow } & 
G_{\tau _{+}\uparrow \tau _{-}\uparrow } & G_{\tau _{+}\uparrow \tau _{-}\downarrow } \\ 
G_{\tau _{+}\downarrow \tau _{+}\uparrow } & G_{\tau _{+}\downarrow \tau _{+}\downarrow } & 
G_{\tau _{+}\downarrow \tau _{-}\uparrow } & G_{\tau _{+}\downarrow \tau _{-}\downarrow } \\ 
G_{\tau _{-}\uparrow \tau _{+}\uparrow } & G_{\tau _{-}\uparrow \tau _{+}\downarrow } & 
G_{\tau _{-}\uparrow \tau _{-}\uparrow } & G_{\tau _{-}\uparrow \tau _{-}\downarrow } \\ 
G_{\tau _{-}\downarrow \tau _{+}\uparrow } & G_{\tau _{-}\downarrow \tau _{+}\downarrow } & 
G_{\tau _{-}\downarrow \tau _{-}\uparrow } & G_{\tau _{-}\downarrow \tau _{-}\downarrow }
\end{array}
\right) \nonumber\\
&=&
\frac{1}{2\pi }e^{-\frac{x^{2}+y^{2}+x^{\prime 2}+y^{\prime 2}}{2}}
\left( 
\begin{array}{cc}
G_{11} & 0 \\ 
0 & G_{22}
\end{array}
\right)
\end{eqnarray}
with
\begin{eqnarray}
G_{11}=
\left( 
\begin{array}{cc}
\frac{1}{(\varepsilon+{\rm i}0^{+})+k_{z}} & \frac{\left( -{\rm i}-1\right) }{\sqrt{2}} \frac{1}{(\varepsilon+{\rm i}0^{+})+k_{z}}  \\ 
\frac{\left( {\rm i}-1\right) }{\sqrt{2}} \frac{1}{(\varepsilon+{\rm i}0^{+})+k_{z}} & \frac{1}{(\varepsilon+{\rm i}0^{+})+k_{z}}
\end{array}
\right)
\end{eqnarray}
\begin{eqnarray}
G_{22}=
\left( 
\begin{array}{cc} 
\frac{1}{(\varepsilon+{\rm i}0^{+})-k_{z}} & \frac{\left({\rm i}-1\right) }{\sqrt{2}} \frac{1}{(\varepsilon+{\rm i}0^{+})-k_{z}} \\ 
\frac{\left( -{\rm i}-1\right) }{\sqrt{2}} \frac{1}{(\varepsilon+{\rm i}0^{+})-k_{z}} & \frac{1}{(\varepsilon+{\rm i}0^{+})-k_{z}}
\end{array}
\right).
\end{eqnarray}
One intriguing observation is that the propagator $G_{11}$ possesses a momentum of $-k_{z}$ and a SU(2) structure of $(-\sigma_{x}+\sigma_{y})$, while the propagator $G_{22}$ contains $k_{z}$ and $(-\sigma_{x}-\sigma_{y})$, potentially due to the spin-momentum locking of the helical hinge states. 

After the Fourier transformation, we can get the real-space Green's function 
\begin{eqnarray}\label{eq:rG}
G\left( z,\varepsilon \right) = \frac{{\rm i}}{4\pi} e^{-\frac{\left( x^{2}+y^{2}+x^{\prime 2}+y^{\prime 2}\right) }{2}}
\left( 
\begin{array}{cc}
g_{11}& 0 \\ 
0 & g_{22}
\end{array}
\right)
\end{eqnarray}
with 
\begin{eqnarray}
g_{11} =e^{-{\rm i}\varepsilon z}
\left( 
\begin{array}{cc}
\left( \mathrm{sgn}\left[ z\right] -1\right)  
& \left( \mathrm{sgn}\left[ z\right] -1\right)\frac{\left( -{\rm i}-1\right) }{\sqrt{2}}\\ 
\left( \mathrm{sgn}\left[ z\right] -1\right) \frac{\left({\rm i}-1\right) }{\sqrt{2}} 
& \left( \mathrm{sgn}\left[ z\right]-1\right) 
\end{array}
\right),
\end{eqnarray}
\begin{eqnarray}
g_{22}=e^{{\rm i}\varepsilon z}
\left( 
\begin{array}{cc} 
\left( \mathrm{sgn}\left[ z\right] +1\right)  
&\left( \mathrm{sgn}\left[ z\right] +1\right) \frac{\left({\rm i}-1\right) }{\sqrt{2}} \\ 
\left( \mathrm{sgn}\left[ z\right] +1\right) \frac{\left(-{\rm i}-1\right) }{\sqrt{2}} 
& \left( \mathrm{sgn}\left[ z\right]+1\right) 
\end{array}
\right)
\end{eqnarray}
where $\mathrm{sgn}\left[ z\right]$ is the sign function. We can see that $g_{11(22)}=0$ when $z>0 (z<0)$, indicating the Green's function depends on the $f (d)$ orbitals. This peculiar dependence on the sign of $z$ implies that the helicity of the Green's functions, \textit{i.e.}, the propagator is direction-dependent.

Then we can arrive the RKKY interaction of HOTI in the helical hinge
\begin{eqnarray}\label{eq:results}
H_{1,2}^{RKKY}&=&F_{1}(z,\varepsilon_{F})\left(\bm{S}_{1}\cdot\bm{S}_{2}+S_{1x}S_{2x}-S_{1y}S_{2y}\right)\notag\\
&&+F_{2}(z,\varepsilon_{F})\left( \bm{S}_{1}\bm{\times S}_{2}\right) _{y}
\end{eqnarray}
with 
\begin{eqnarray}\label{eq:rangefunctions}
F_{1}(z,\varepsilon_{F})=
\frac{J^{2}}{4\pi ^{3}|z|}
\left[ \cos \left(2z\varepsilon _{F}\right)-1\right] 
e^{-\left(x^{2}+y^{2}+x^{\prime 2}+y^{\prime 2}\right) }\\
F_{2}(z,\varepsilon_{F})=-\frac{\sqrt{2}J^{2}}{4\pi ^{3}|z|}
\sin \left(2z\varepsilon _{F}\right)
e^{-\left( x^{2}+y^{2}+x^{\prime 2}+y^{\prime 2}\right) }.
\end{eqnarray}
by combining the Eq.~(\ref{eq:intH}), Eq.~(\ref{eq:RKKY}) and Eq.~(\ref{eq:rG}) and introducing a cutoff function $ \exp(-\varepsilon ^{2}/\Lambda ^{2})$ with $\Lambda \rightarrow 0 $. We can turn the analytical expressions to the International System of Units for facilitating the experimentalists: 
\begin{eqnarray}
F_{1}(z,\varepsilon_{F})=
\frac{am_{0}\hbar ^{-2}J^{2}}{4\pi ^{3}|z|}
\left[ \cos \left(2am_{0}\hbar^{-2}z\varepsilon _{F}\right)-1\right] \notag\\
\times e^{-a^{-2}\left(x^{2}+y^{2}+x^{\prime 2}+y^{\prime 2}\right) }\\
F_{2}(z,\varepsilon_{F})=
-\frac{\sqrt{2}am_{0}\hbar ^{-2}J^{2}}{4\pi ^{3}|z|}
\sin \left(2am_{0}\hbar ^{-2}z\varepsilon _{F}\right) \notag\\
\times e^{-a^{-2}\left( x^{2}+y^{2}+x^{\prime 2}+y^{\prime 2}\right) }
\end{eqnarray}
$\hbar$ is the reduced Planck constant, and $a$ is half crystal constant of SnTe about 0.316 $nm$ \cite{schindler2018higher, schindler2022topological}. We take spin-spin coupling constant $J=10$ $meVnm$.
It should be noted that there exist two non-equivalent branches ($z>0$ and $z<0$), and the only factor that sets them apart for a given inter-impurity distance is the sign of the DM term, indicating the helicity type. Here Fermi energy $ \varepsilon _{F}>0 (<0) $ corresponds to n(p)-type doping.  The different type of doping cause a sign reversal of the DM interaction $ \left( \bm{S}_{1}\bm{\times S}_{2}\right) _{y} $. The long-range asymptotic behavior of the RKKY interaction is 
\begin{eqnarray}
H_{1,2}^{RKKY}(z\rightarrow\infty)&\simeq & \frac{1}{z}\left[\bm{S}_{1}\cdot\bm{S}_{2}+S_{1x}S_{2x}-S_{1y}S_{2y}\right.\notag\\
&&\left. -\left(\bm{S}_{1}\bm{\times S}_{2}\right) _{y}\right].
\end{eqnarray}

\begin{figure}[thp]
\begin{center}
\includegraphics[width=\columnwidth]{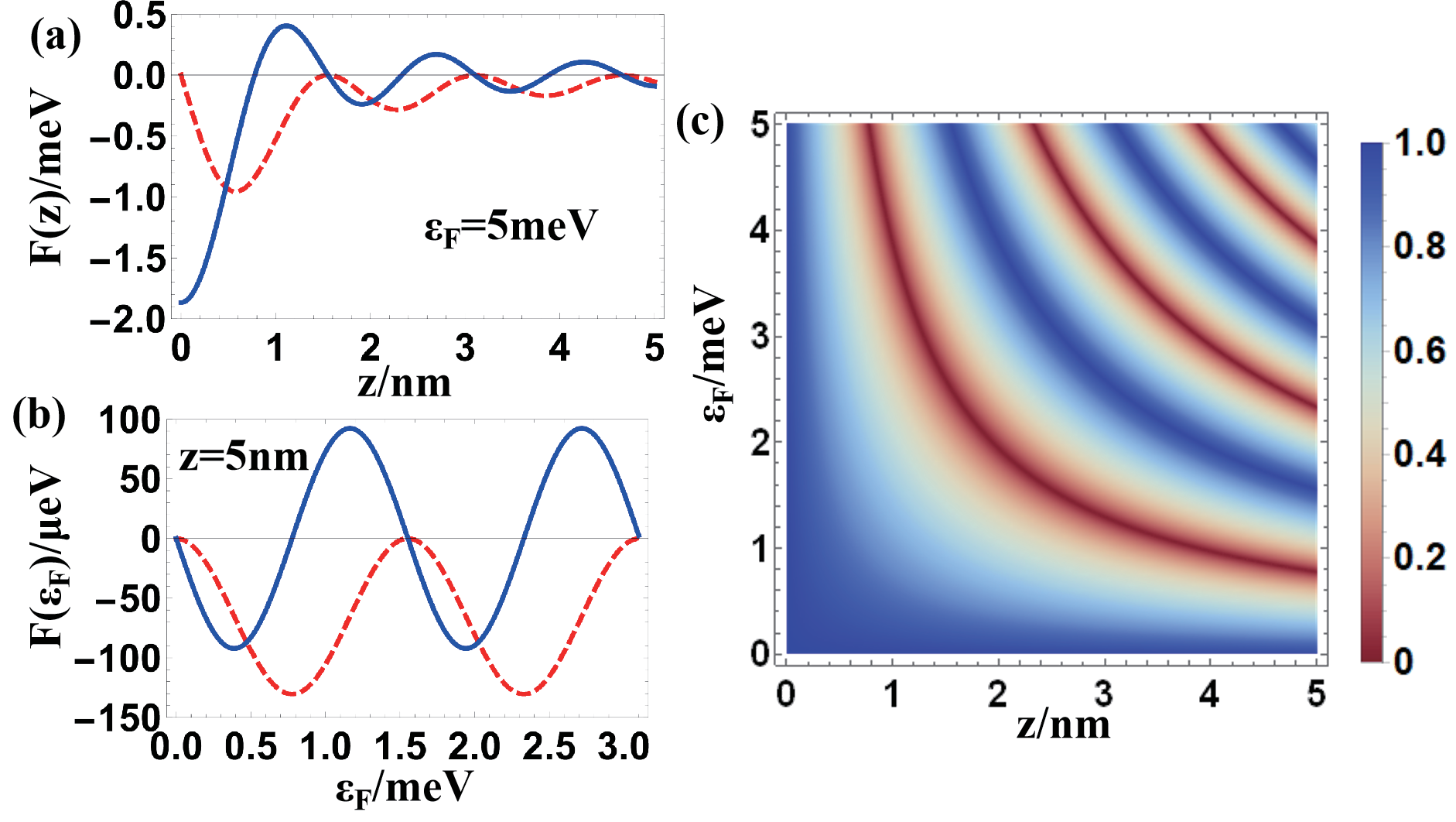} \\
\parbox[c]{\columnwidth}{\caption{
Helical HOTIs' range functions depending on (a) the impurity spacing $z$, (b) the Fermi energy $\varepsilon _{F}$ , and (c) the proportion of DM interaction. (a) The range functions oscillating decay with impurity spacing $z$ at  $\varepsilon _{F}=5$ meV. (b) The range functions depending on the Fermi energy $\varepsilon _{F}$ oscillate with the same period as simple sine functions corresponding to impurity spacing $z=5$ nm. The red solid line stands for the collinear terms $F_{1}$ and the blue dashed line stands for the non-collinear term (DM interaction)$F_{2}$. (c) The proportion of DM interaction in the total RKKY interaction $|F_{2}|/(|F_{1}|+|F_{2}|)$ depending on the impurity spacing $z$ and the Fermi energy $\varepsilon _{F}$. We take coupling strength $J = 10$ $\mathrm{meV}\cdot\mathrm{nm}$.}\label{fig:helicalZE}}
\end{center}
\end{figure}

The range functions depend on the impurity spacing $z$, the Fermi energy $\varepsilon _{F}$ and the distribution of carriers in $xy$ plane. In Fig.~\ref{fig:helicalZE}(a), we can see the range functions $F_{1}(z,\varepsilon_{F}),F_{2}(z,\varepsilon_{F})$ depend on the impurity spacing $z$, showing the same damped oscillatory behavior. The hinge RKKY interaction decay as $z^{-1}$, which is in line with the previous results of the RKKY interaction in helical edges of topological superconductors\cite{laubscher2023rkky}. The 1D topological higher-order boundary could slow down the decay of the RKKY interaction from $1/z^2$ of the 2D common topological surface\cite{yarmohammadi2023noncollinear,yarmohammadi2020effective,zhu2011electrically} to $1/z$, and hence drastically enhances its magnitude. This stronger RKKY interaction in 1D nontrivial helical hinge is more promising for potential applications in topological spintronics. In helical HOTIs, the Heisenberg and Ising type interaction are always ferromagnetic while DM type interaction shows anisotropic gyromagnetism. With increasing Fermi energy $\varepsilon _{F}$, the range functions of helical HOTIs show the simple sine function oscillating behavior with the same period, shown in Fig.~\ref{fig:helicalZE}(b). If we finely tune the parameters $z$ and $\varepsilon _{F}$, we can get different spin interactions within different regimes of the parameters. For example, the RKKY interaction reduces to a pure DM interaction when the inter-impurity distance and the Fermi energy are rather small, and the RKKY interaction reduces to a anisotropic Heisenberg interaction when the inter-impurity distance and the Fermi energy reach the critical values $z\varepsilon _{F}=(n+1/2)\pi$. Fig.~\ref{fig:helicalZE}(c) shows the proportion of DM interaction in the total RKKY interaction. We can see that the proportion oscillate with increasing impurity spacing $z$ and the Fermi energy $\varepsilon _{F}$, and we can finely tune the hinge RKKY interaction between the collinear anisotropic Heisenberg interaction and the non-collinear DM interaction. 

\begin{figure}[thp]
\begin{center}
\includegraphics[width=\columnwidth]{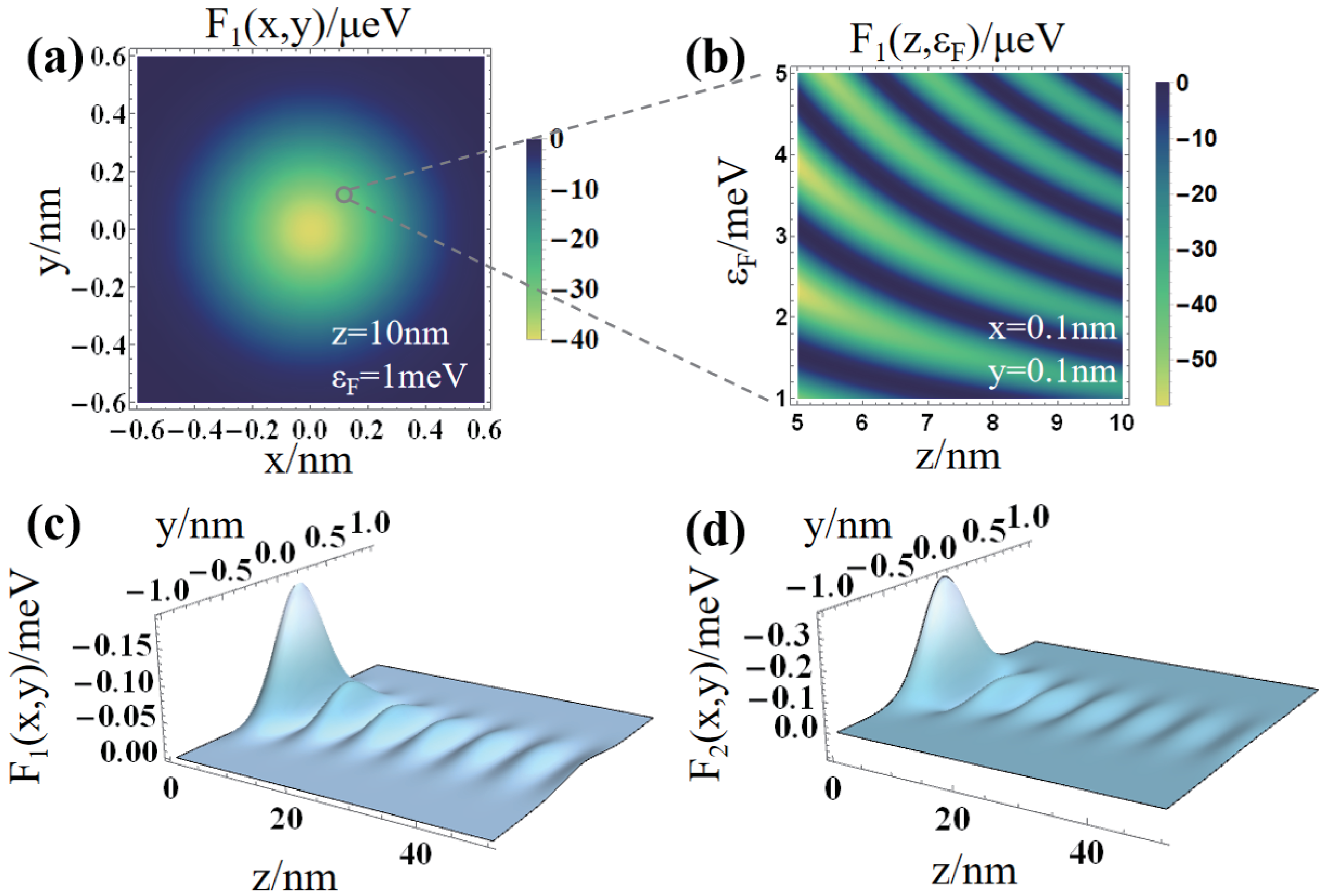}\\
\parbox[c]{\columnwidth}{\caption{
(a)The isotropic and exponential decayed distribution of range function $F_{1}$ in the $xy$ plane corresponding to $z=10$ nm and $\varepsilon _{F}=1$ meV. (b) The range function $F_{1}$ off the hinge at $x=0.1, y=0.1$ nm depending on the impurity spacing $z$ and the Fermi energy $\varepsilon _{F}$. (c) and (d) show the distribution of the range function $F_{1}$ and $F_{2}$ in $yz$ plane at $x=0, \varepsilon _{F}=1$ meV.}\label{fig:helicalXYZ}}
\end{center}
\end{figure}

The distribution of the range function $F_{1}$ in the $xy$ plane depending on inter-impurity distance $z$ and the Fermi energy $\varepsilon _{F}$ are shown in Fig.~\ref{fig:helicalXYZ}(a). The range functions are isotropic in the $xy$ plane and decrease exponentially, which is a natural consequence resulting from the trivial bulk states. The range function $F_2$ shows exactly the same behavior as $F_1$. When the impurities locate off the hinge, for instance, $x=0.1$ nm, $y=0.1$ nm, the RKKY interaction show almost the same behavior as in the hinge, except for a decrease in magnitude, as shown in Fig.~\ref{fig:helicalXYZ}(b). The robust nature of the hinge RKKY interaction guarantees the durability of the spintronic devices that may be realized in the helical hinge of HOTIs. We also show the distribution of range functions $F_{1}$ and $F_{2}$ depending on coordinates $y$ and $z$ in Fig.~\ref{fig:helicalXYZ}(c) and (d), which clearly indicate an oscillatory decay pattern along the $z$ axis and an exponential decay along the $y$ axis.

\section{RKKY interaction in the Dirac-type bulk states} 
For bulk doping, density functional theory calculations\cite{PhysRevB.84.245418,PhysRevLett.108.206801,PhysRevLett.109.266405} and quantum Monte Carlo calculation\cite{PhysRevB.89.115101} find a complicated anisotropic spin texture in magnetically doped TIs. Therefore, the 3D Dirac-type bulk states attribute to the RKKY interaction in the doped regimes, and may also induce the complicated RKKY interaction due to the profound relation between the bulk states and the hinge states. This relation is reflected by the fact that the Hamiltonian of hinge states is derived from the Dirac-type bulk Hamiltonian of HOTIs. Here we start from the 3D Dirac-type bulk Hamiltonian\cite{schindler2018higher}
\begin{eqnarray}
{H_b (\boldsymbol{k}) = k_x \tau_x \sigma_x + k_y \tau_x\sigma_y + k_z \tau_x \sigma_z},
\end{eqnarray}
where the Pauli matrixes $ \tau,\sigma $ are defined as before, and examine the bulk RKKY interaction in the helical HOTIs. The Green's function in momentum space can be calculated directly as
\begin{eqnarray}
G(\boldsymbol{k},\varepsilon)=\frac{\varepsilon\tau_0\sigma_0+\tau_x\boldsymbol{k}\cdot\bm{\sigma}}{\varepsilon^2 -k^2}. 
\end{eqnarray}
And by applying Fourier transform on $ G(\boldsymbol{k},\varepsilon) $, we can arrive the Green's function in real space
\begin{eqnarray}
G(\boldsymbol{R},\varepsilon)=\frac{1}{(2\pi)^3}\int G(\boldsymbol{k},\varepsilon)e^{{\rm i}\boldsymbol{k}\cdot\boldsymbol{R}}{\rm d}^3 \boldsymbol{k}. 
\end{eqnarray}
We can replace $ \boldsymbol{k} $ with $ \boldsymbol{k}_{\parallel}+\boldsymbol{k}_{\bot} $, where $ \boldsymbol{k}_{\parallel}$ is the part parallel to $ \boldsymbol{R} $ and $\boldsymbol{k}_{\bot}$ the perpendicular part. For $\boldsymbol{k}_{\bot}\cdot\boldsymbol{R}= 0$, we have 
\begin{eqnarray}
G(\boldsymbol{R},\varepsilon)&=&\frac{1}{(2 \pi)^3} \int \frac{\varepsilon \tau_0 \sigma_0 + \tau_x\left( \boldsymbol{k}_{\parallel} +\boldsymbol{k}_{\bot} \right) \cdot\bm{\sigma}}{\varepsilon^2 - k^2} e^{{\rm i} \boldsymbol{k}_{\parallel}\cdot \boldsymbol{R}}{\rm d}^3 \boldsymbol{k}\nonumber\\
&=&\frac{- \varepsilon e^{{\rm i} R \varepsilon}}{4 \pi R} \tau_0 \sigma_0+\left( \frac{e^{{\rm i} R \varepsilon}}{4 \pi {\rm i} R^2} - \frac{\varepsilon e^{{\rm i} R \varepsilon}}{4 \pi R} \right) \tau_x \hat{\boldsymbol{R}} \cdot\bm{\sigma}.
\end{eqnarray}
Similarly, 
\begin{eqnarray}
G(\boldsymbol{-R},\varepsilon)=\frac{- \varepsilon e^{{\rm i} R \varepsilon}}{4 \pi R} \tau_0 \sigma_0-\left( \frac{e^{{\rm i} R \varepsilon}}{4 \pi {\rm i} R^2} - \frac{\varepsilon e^{{\rm i} R \varepsilon}}{4 \pi R} \right) \tau_x \hat{\boldsymbol{R}} \cdot\bm{\sigma} .
\end{eqnarray}

The interaction Hamiltonian is 
\begin{eqnarray}
H_{i}^{\mathrm{int}}=J (\tau_0+\tau_x) \otimes ( \bm{\sigma} \cdot \bm{S}_{i})\delta(\boldsymbol{R}-\boldsymbol{R}_{i}).
\end{eqnarray}
After a standard derivation of the RKKY interaction by the integral 
\begin{eqnarray}
H_{bulk 1,2}^{\mathrm{RKKY}}=-\frac{1}{\pi } \mathrm{Im}\int_{-\infty }^{\varepsilon_{F}}\mathrm{Tr}\left[H_{1}^{int}G_{1}(\boldsymbol{R},\varepsilon)H_{2}^{int} G_{2}(-\boldsymbol{R},\varepsilon)\right] d\varepsilon,
\end{eqnarray}
we finally get the bulk RKKY interaction depending on the orientation of two impurities
\begin{eqnarray}
H^{RKKY}_{bulk 1,2}&=&f_{1}(R,\varepsilon_F)\bm{S}_{1}\cdot\bm{S}_{2}
+f_{2}(R,\varepsilon_F)\bm{S}_{1}\cdot\bm{M}\cdot\bm{S}_{2}\notag\\
&&+f_{3}(R, \varepsilon_F)\bm{n}\cdot(\bm{S}_{1}\times\bm{S}_{2})
\end{eqnarray}
with
\begin{eqnarray}
f_1(R, \varepsilon_F)&=&\frac{J^2}{8\pi^3 R^5}[-\cos{(2R\varepsilon_F)} -2 R\varepsilon_F \sin{(2R\varepsilon_F)}\notag\\
&&+2 R^{2}\varepsilon_F^{2} \cos{(2R\varepsilon_F)}],\\ 
f_2(R, \varepsilon_F)&=&\frac{J^2}{16\pi^3 R^5}[-5\cos{(2R\varepsilon_F)}-6 R\varepsilon_F \sin{(2R\varepsilon_F)}\notag\\
&& +2 R^{2}\varepsilon_F^{2} \cos{(2R\varepsilon_F)}],\\
f_3(R, \varepsilon_F)&=&\frac{J^2}{2\pi^3 R^5}[-\sin{(2R\varepsilon_F)}+2 R\varepsilon_F \cos{(2R\varepsilon_F)}\notag\\
&& +R^{2}\varepsilon_F^{2}\sin{(2R\varepsilon_F)}].
\end{eqnarray}
The corresponding International System of Units expressions are omitted here for brevity.
The second-order tensor of rotation
\begin{widetext} 
\begin{eqnarray}
\bm{M}=\left(
\begin{array}{ccc} 
1+\cos{2\theta}-2\sin^{2}{\theta}\cos{2\varphi} &-2\sin^{2}{\theta}\sin{2\varphi} &-2\sin{2\theta}\cos{\varphi}\\
-2\sin^{2}{\theta}\sin{2\varphi} &1+\cos{2\theta}+2\sin^{2}{\theta}\cos{2\varphi} &-2\sin{2\theta}\sin{\varphi}\\
-2\sin{2\theta}\cos{\varphi} &-2\sin{2\theta}\sin{\varphi} &\sin^{2}{\theta}\sin{2\varphi}
\end{array}
\right),
\end{eqnarray}
\end{widetext}
and the orientation vector $\bm{n}=(\cos\varphi\sin\theta,\sin\varphi\sin\theta,\cos\theta)$. Here the angle $\theta$ is defined between $\boldsymbol{R}$ and $z$ axis, and the angle $\varphi$ is defined between the projection of $\boldsymbol{R}$ on $xy$ plane and the $x$ axis. The bulk RKKY interaction has complicated formalism consist of a Heisenberg term, a twisted Ising term, and a DM term along the orientation vector, similar to the RKKY interaction in TI, topological semimetal and systems with spin-orbit coupling \cite{yarmohammadi2023noncollinear,chang2015rkky,duan2022prolonged,PhysRevB.101.075421,PhysRevB.80.155302,duan2018bulk}.

The long-range asymptotic behavior of the RKKY interaction is 
\begin{eqnarray}
H^{RKKY}_{bulk 1,2}(R\rightarrow\infty)&\simeq & \frac{1}{R^3}\left[\bm{S}_{1}\cdot\bm{S}_{2}+\bm{S}_{1}\cdot\bm{M}\cdot\bm{S}_{2}\right. \notag\\
 &&\left. +\bm{n}\cdot(\bm{S}_{1}\times \bm{S}_{2})\right].
\end{eqnarray} 
Notice that the range function monotonically decreases as $1/R^3$ in this case, the same as the conventional RKKY interaction.

\begin{figure}[thp]
\begin{center}
\includegraphics[width=\columnwidth]{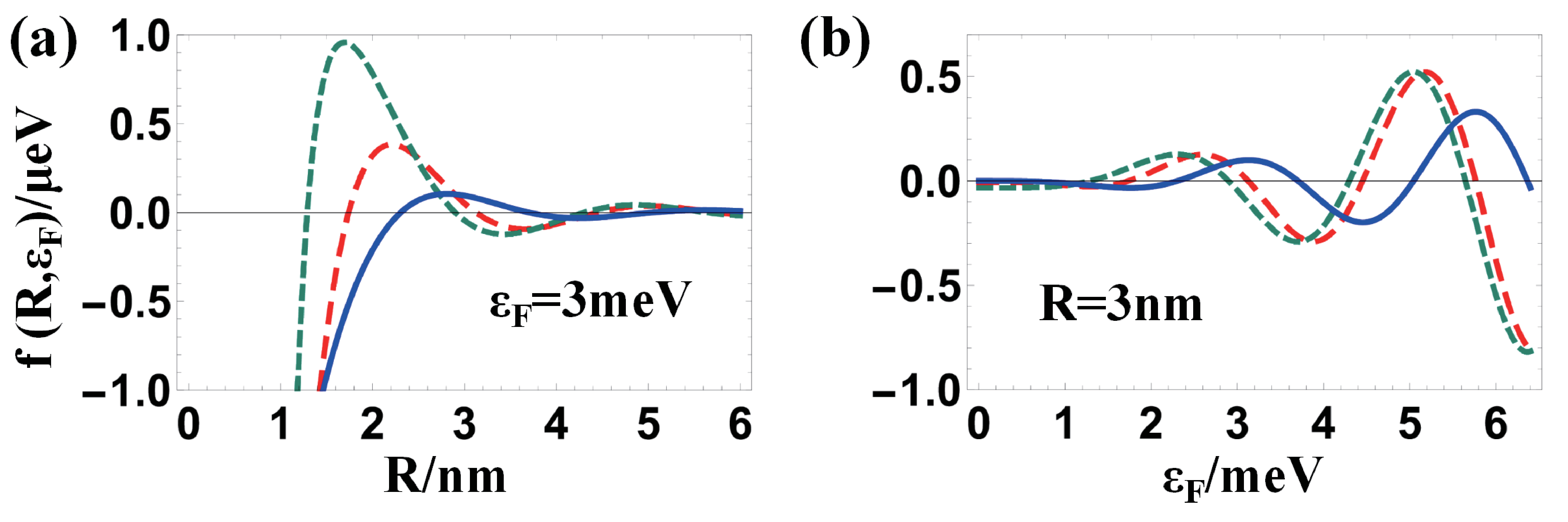}
\parbox[c]{\columnwidth}{\caption{
(a)The bulk's range functions $f_{1}(R, \varepsilon_F)$,$f_{2}(R, \varepsilon_F)$,$f_{3}(R, \varepsilon_F)$ depending on impurity spacing $R$ at Fermi energy $\varepsilon _{F}=3$ meV denoted by red (long-dashed), green (short-dashed), blue (solid) line respectively. These range functions oscillate with the same period and decay with $R^{-3}$ asymptotically.(b) The bulk's range functions depending on Fermi energy $\varepsilon _{F}$ at impurity spacing $R=3$ nm.}\label{fig:Dirac}}
\end{center}
\end{figure}

The range functions of bulk RKKY interaction are showed in Fig.~\ref{fig:Dirac} and depend on impurity spacing $R$ and Fermi energy $\varepsilon _{F}$. We can see that the bulk RKKY interaction decay much faster ($R^{-3}$) than the hinge RKKY interaction ($R^{-1}$) in Fig.~\ref{fig:Dirac}(a). The range functions show the conventional oscillating decay behavior as in normal metals and semiconductors. The primary distinction between the bulk and hinge RKKY interactions is the different dependence on the Fermi energy $\varepsilon _{F}$. The range functions of bulk RKKY interaction exhibit greater oscillation with increasing Fermi energy. All these range functions oscillate with the same period.

\begin{figure}[thp]
\begin{center}
\includegraphics[width=\columnwidth]{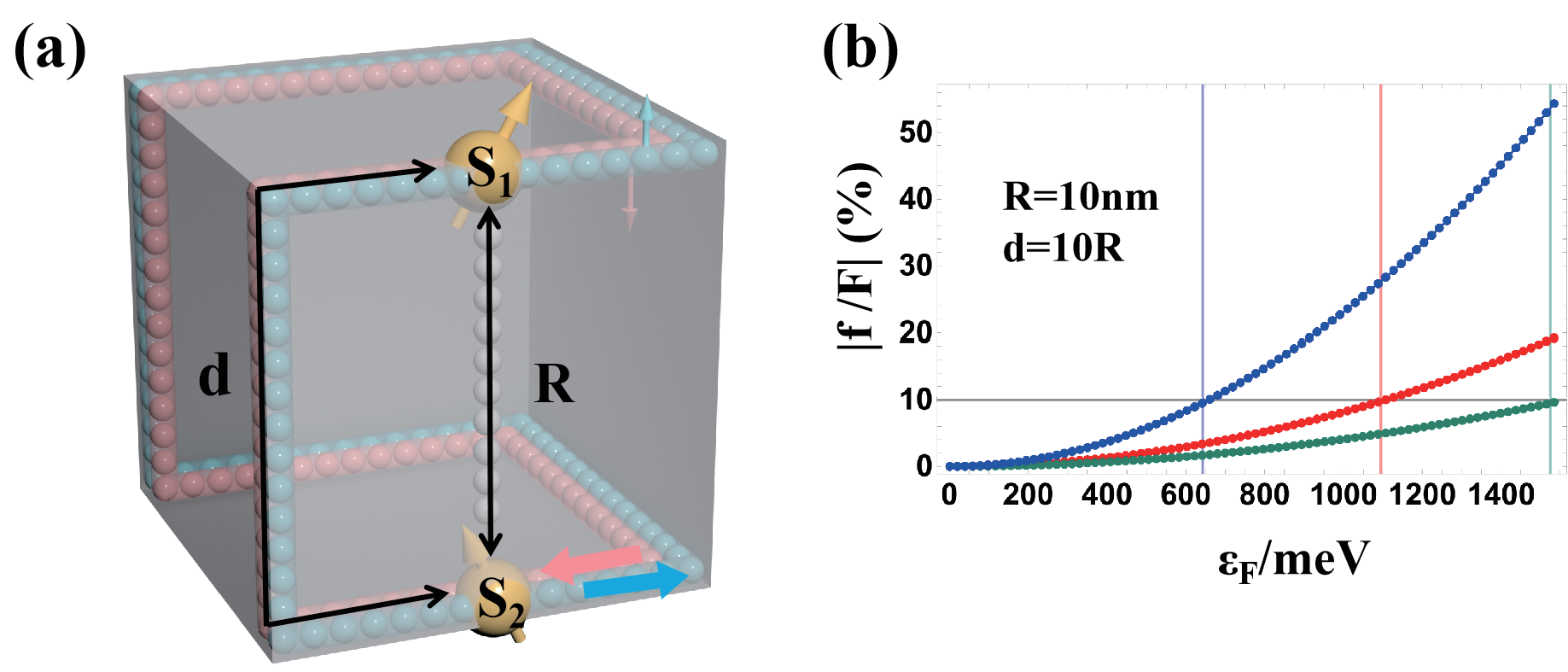}
\parbox[c]{\columnwidth}{\caption{
(a)The schematic diagram of RKKY interaction mediated by both hinge and bulk carriers between impurities in complex configuration. The distance corresponding to bulk carriers is specified as $R$, while for hinge carriers it is indicated as $d$. We take the hinge spacing $R=1$nm and the bulk spacing $d=10R$. (b)The percentage of bulk to hinge correction depending on the Fermi energy $\varepsilon _{F}$ at $R=$10 nm,$d=10$ $R$. The red, green and blue lines stand for the Heisenberg, (twisted) Ising and DM term, respectively. Here the value of range functions are represented by the minimum in every period.}\label{fig:hingeAndBulk}}
\end{center}
\end{figure}

The RKKY interaction can be mediated by both the helical itinerant carriers and the Dirac bulk carriers in the doped regime, as shown in Fig.~\ref{fig:hingeAndBulk}(a). At appropriate parameter selection, the contribution of hinge and bulk to RKKY interaction are comparable. We show the correction of the bulk RKKY interaction to the hinge RKKY interaction when the Fermi energy increases, shown in Fig.~\ref{fig:hingeAndBulk}(b).  The proportion $ f/F $ change with $\varepsilon_{F}^{2}$. The contribution of the bulk RKKY interaction more than 10\% of the hinge RKKY interaction when the Fermi energy is about 1100 meV for the Heisenberg term, 1510 meV for the (twisted) Ising term and 620 meV for the DM term. By adjusting the Fermi energy, we can utilize this phenomena to effectively transform the RKKY interaction into a dominant DM term within the spin interaction formalism.

\section{Conclusions}
In conclusion, our theoretical exploration of the RKKY interaction within helical HOTIs unveils distinct interaction mechanisms mediated by hinge and Dirac-type bulk carriers. Specifically, we discovered that the hinge-mediated RKKY interaction encompasses a comprehensive suite of terms: the Heisenberg, $x$-Ising, $y$-Ising, and DM terms, with the latter oriented along the $y$ direction. Notably, these interactions exhibit a decay proportional to $z^{-1}$ with increasing impurity spacing, and oscillate in strength according to sine functions as a function of the Fermi energy, $\varepsilon_F$. This behavior underscores a consistent ferromagnetic coupling in both Heisenberg-type and Ising-type interactions, while DM-type interactions demonstrate alternating behaviors. Moreover, the isotropic decay of these interactions in the $xy$ plane further delineates the nuanced role of hinge carriers. Conversely, interactions mediated by bulk carriers are characterized by Heisenberg, twisted Ising, and DM terms, showcasing a $R^{-3}$ decay and conventional oscillatory decay with increased impurity spacing. The interplay between hinge and bulk interactions presents a fertile ground for developing advanced quantum devices, leveraging the unique properties of helical HOTIs. Our findings not only deepen the understanding of RKKY interactions in these novel materials but also open avenues for the design of next-generation spintronics devices, capitalizing on the intricate interplay of hinge and bulk carrier dynamics.

\section*{Acknowledgment}
This work has been supported by the research foundation of Institute for Advanced Sciences of CQUPT (Grant No. E011A2022328).

\bibliography{reference}

\begin{thebibliography}{51}%
\makeatletter
\providecommand \@ifxundefined [1]{%
 \@ifx{#1\undefined}
}%
\providecommand \@ifnum [1]{%
 \ifnum #1\expandafter \@firstoftwo
 \else \expandafter \@secondoftwo
 \fi
}%
\providecommand \@ifx [1]{%
 \ifx #1\expandafter \@firstoftwo
 \else \expandafter \@secondoftwo
 \fi
}%
\providecommand \natexlab [1]{#1}%
\providecommand \enquote  [1]{``#1''}%
\providecommand \bibnamefont  [1]{#1}%
\providecommand \bibfnamefont [1]{#1}%
\providecommand \citenamefont [1]{#1}%
\providecommand \href@noop [0]{\@secondoftwo}%
\providecommand \href [0]{\begingroup \@sanitize@url \@href}%
\providecommand \@href[1]{\@@startlink{#1}\@@href}%
\providecommand \@@href[1]{\endgroup#1\@@endlink}%
\providecommand \@sanitize@url [0]{\catcode `\\12\catcode `\$12\catcode
  `\&12\catcode `\#12\catcode `\^12\catcode `\_12\catcode `\%12\relax}%
\providecommand \@@startlink[1]{}%
\providecommand \@@endlink[0]{}%
\providecommand \url  [0]{\begingroup\@sanitize@url \@url }%
\providecommand \@url [1]{\endgroup\@href {#1}{\urlprefix }}%
\providecommand \urlprefix  [0]{URL }%
\providecommand \Eprint [0]{\href }%
\providecommand \doibase [0]{https://doi.org/}%
\providecommand \selectlanguage [0]{\@gobble}%
\providecommand \bibinfo  [0]{\@secondoftwo}%
\providecommand \bibfield  [0]{\@secondoftwo}%
\providecommand \translation [1]{[#1]}%
\providecommand \BibitemOpen [0]{}%
\providecommand \bibitemStop [0]{}%
\providecommand \bibitemNoStop [0]{.\EOS\space}%
\providecommand \EOS [0]{\spacefactor3000\relax}%
\providecommand \BibitemShut  [1]{\csname bibitem#1\endcsname}%
\let\auto@bib@innerbib\@empty
\bibitem [{\citenamefont {Kim}\ \emph {et~al.}(2020)\citenamefont {Kim},
  \citenamefont {Jacob},\ and\ \citenamefont {Rho}}]{kim2020recent}%
  \BibitemOpen
  \bibfield  {author} {\bibinfo {author} {\bibfnamefont {M.}~\bibnamefont
  {Kim}}, \bibinfo {author} {\bibfnamefont {Z.}~\bibnamefont {Jacob}},\ and\
  \bibinfo {author} {\bibfnamefont {J.}~\bibnamefont {Rho}},\ }\bibfield
  {title} {\bibinfo {title} {Recent advances in 2d, 3d and higher-order
  topological photonics},\ }\href {https://doi.org/10.1038/s41377-020-0331-y}
  {\bibfield  {journal} {\bibinfo  {journal} {Light Sci. Appl.}\ }\textbf
  {\bibinfo {volume} {9}},\ \bibinfo {pages} {130} (\bibinfo {year}
  {2020})}\BibitemShut {NoStop}%
\bibitem [{\citenamefont {Xie}\ \emph {et~al.}(2021)\citenamefont {Xie},
  \citenamefont {Wang}, \citenamefont {Zhang}, \citenamefont {Zhan},
  \citenamefont {Jiang}, \citenamefont {Lu},\ and\ \citenamefont
  {Chen}}]{xie2021higher}%
  \BibitemOpen
  \bibfield  {author} {\bibinfo {author} {\bibfnamefont {B.}~\bibnamefont
  {Xie}}, \bibinfo {author} {\bibfnamefont {H.-X.}\ \bibnamefont {Wang}},
  \bibinfo {author} {\bibfnamefont {X.}~\bibnamefont {Zhang}}, \bibinfo
  {author} {\bibfnamefont {P.}~\bibnamefont {Zhan}}, \bibinfo {author}
  {\bibfnamefont {J.-H.}\ \bibnamefont {Jiang}}, \bibinfo {author}
  {\bibfnamefont {M.}~\bibnamefont {Lu}},\ and\ \bibinfo {author}
  {\bibfnamefont {Y.}~\bibnamefont {Chen}},\ }\bibfield  {title} {\bibinfo
  {title} {Higher-order band topology},\ }\href
  {https://doi.org/10.1038/s42254-021-00323-4} {\bibfield  {journal} {\bibinfo
  {journal} {Nat. Rev. Phys.}\ }\textbf {\bibinfo {volume} {3}},\ \bibinfo
  {pages} {520} (\bibinfo {year} {2021})}\BibitemShut {NoStop}%
\bibitem [{\citenamefont {Benalcazar}\ \emph
  {et~al.}(2017{\natexlab{a}})\citenamefont {Benalcazar}, \citenamefont
  {Bernevig},\ and\ \citenamefont {Hughes}}]{benalcazar2017quantized}%
  \BibitemOpen
  \bibfield  {author} {\bibinfo {author} {\bibfnamefont {W.~A.}\ \bibnamefont
  {Benalcazar}}, \bibinfo {author} {\bibfnamefont {B.~A.}\ \bibnamefont
  {Bernevig}},\ and\ \bibinfo {author} {\bibfnamefont {T.~L.}\ \bibnamefont
  {Hughes}},\ }\bibfield  {title} {\bibinfo {title} {Quantized electric
  multipole insulators},\ }\href {https://doi.org/10.1126/science.aah6442}
  {\bibfield  {journal} {\bibinfo  {journal} {Science}\ }\textbf {\bibinfo
  {volume} {357}},\ \bibinfo {pages} {61} (\bibinfo {year}
  {2017}{\natexlab{a}})}\BibitemShut {NoStop}%
\bibitem [{\citenamefont {Liu}\ and\ \citenamefont
  {Wakabayashi}(2017)}]{liu2017novel}%
  \BibitemOpen
  \bibfield  {author} {\bibinfo {author} {\bibfnamefont {F.}~\bibnamefont
  {Liu}}\ and\ \bibinfo {author} {\bibfnamefont {K.}~\bibnamefont
  {Wakabayashi}},\ }\bibfield  {title} {\bibinfo {title} {Novel topological
  phase with a zero berry curvature},\ }\href
  {https://doi.org/10.1103/PhysRevLett.118.076803} {\bibfield  {journal}
  {\bibinfo  {journal} {Phys. Rev. Lett.}\ }\textbf {\bibinfo {volume} {118}},\
  \bibinfo {pages} {076803} (\bibinfo {year} {2017})}\BibitemShut {NoStop}%
\bibitem [{\citenamefont {Lin}\ \emph {et~al.}(2020)\citenamefont {Lin},
  \citenamefont {Wu}, \citenamefont {Wang},\ and\ \citenamefont
  {Jiang}}]{ZhiKangLin74302}%
  \BibitemOpen
  \bibfield  {author} {\bibinfo {author} {\bibfnamefont {Z.-K.}\ \bibnamefont
  {Lin}}, \bibinfo {author} {\bibfnamefont {S.-Q.}\ \bibnamefont {Wu}},
  \bibinfo {author} {\bibfnamefont {H.-X.}\ \bibnamefont {Wang}},\ and\
  \bibinfo {author} {\bibfnamefont {J.-H.}\ \bibnamefont {Jiang}},\ }\bibfield
  {title} {\bibinfo {title} {Higher-order topological spin hall effect of
  sound},\ }\href {https://doi.org/10.1088/0256-307X/37/7/074302} {\bibfield
  {journal} {\bibinfo  {journal} {Chinese Physics Letters}\ }\textbf {\bibinfo
  {volume} {37}},\ \bibinfo {eid} {074302} (\bibinfo {year}
  {2020})}\BibitemShut {NoStop}%
\bibitem [{\citenamefont {Chen}\ \emph {et~al.}(2024)\citenamefont {Chen},
  \citenamefont {Liu}, \citenamefont {Chen},\ and\ \citenamefont
  {Zhou}}]{HuanChen17202}%
  \BibitemOpen
  \bibfield  {author} {\bibinfo {author} {\bibfnamefont {H.}~\bibnamefont
  {Chen}}, \bibinfo {author} {\bibfnamefont {Z.-R.}\ \bibnamefont {Liu}},
  \bibinfo {author} {\bibfnamefont {R.}~\bibnamefont {Chen}},\ and\ \bibinfo
  {author} {\bibfnamefont {B.}~\bibnamefont {Zhou}},\ }\bibfield  {title}
  {\bibinfo {title} {Higher-order topological anderson insulator on the
  sierpi{\' n}ski lattice},\ }\href {https://doi.org/10.1088/1674-1056/ad09d4}
  {\bibfield  {journal} {\bibinfo  {journal} {Chinese Physics B}\ }\textbf
  {\bibinfo {volume} {33}},\ \bibinfo {eid} {017202} (\bibinfo {year}
  {2024})}\BibitemShut {NoStop}%
\bibitem [{\citenamefont {Ni}\ \emph {et~al.}(2020)\citenamefont {Ni},
  \citenamefont {Li}, \citenamefont {Weiner}, \citenamefont {Al{\`u}},\ and\
  \citenamefont {Khanikaev}}]{ni2020demonstration}%
  \BibitemOpen
  \bibfield  {author} {\bibinfo {author} {\bibfnamefont {X.}~\bibnamefont
  {Ni}}, \bibinfo {author} {\bibfnamefont {M.}~\bibnamefont {Li}}, \bibinfo
  {author} {\bibfnamefont {M.}~\bibnamefont {Weiner}}, \bibinfo {author}
  {\bibfnamefont {A.}~\bibnamefont {Al{\`u}}},\ and\ \bibinfo {author}
  {\bibfnamefont {A.~B.}\ \bibnamefont {Khanikaev}},\ }\bibfield  {title}
  {\bibinfo {title} {Demonstration of a quantized acoustic octupole topological
  insulator},\ }\href {https://doi.org/10.1038/s41467-020-15705-y} {\bibfield
  {journal} {\bibinfo  {journal} {Nat. Commun.}\ }\textbf {\bibinfo {volume}
  {11}},\ \bibinfo {pages} {2108} (\bibinfo {year} {2020})}\BibitemShut
  {NoStop}%
\bibitem [{\citenamefont {Liu}\ \emph {et~al.}(2020)\citenamefont {Liu},
  \citenamefont {Ma}, \citenamefont {Zhang}, \citenamefont {Zhang},
  \citenamefont {Yang}, \citenamefont {You}, \citenamefont {Gao}, \citenamefont
  {Xiang}, \citenamefont {Cui},\ and\ \citenamefont {Zhang}}]{liu2020octupole}%
  \BibitemOpen
  \bibfield  {author} {\bibinfo {author} {\bibfnamefont {S.}~\bibnamefont
  {Liu}}, \bibinfo {author} {\bibfnamefont {S.}~\bibnamefont {Ma}}, \bibinfo
  {author} {\bibfnamefont {Q.}~\bibnamefont {Zhang}}, \bibinfo {author}
  {\bibfnamefont {L.}~\bibnamefont {Zhang}}, \bibinfo {author} {\bibfnamefont
  {C.}~\bibnamefont {Yang}}, \bibinfo {author} {\bibfnamefont {O.}~\bibnamefont
  {You}}, \bibinfo {author} {\bibfnamefont {W.}~\bibnamefont {Gao}}, \bibinfo
  {author} {\bibfnamefont {Y.}~\bibnamefont {Xiang}}, \bibinfo {author}
  {\bibfnamefont {T.~J.}\ \bibnamefont {Cui}},\ and\ \bibinfo {author}
  {\bibfnamefont {S.}~\bibnamefont {Zhang}},\ }\bibfield  {title} {\bibinfo
  {title} {Octupole corner state in a three-dimensional topological circuit},\
  }\href {https://doi.org/10.1038/s41377-020-00381-w} {\bibfield  {journal}
  {\bibinfo  {journal} {Light Sci. Appl.}\ }\textbf {\bibinfo {volume} {9}},\
  \bibinfo {pages} {145} (\bibinfo {year} {2020})}\BibitemShut {NoStop}%
\bibitem [{\citenamefont {Hackenbroich}\ \emph {et~al.}(2021)\citenamefont
  {Hackenbroich}, \citenamefont {Hudomal}, \citenamefont {Schuch},
  \citenamefont {Bernevig},\ and\ \citenamefont
  {Regnault}}]{hackenbroich2021fractional}%
  \BibitemOpen
  \bibfield  {author} {\bibinfo {author} {\bibfnamefont {A.}~\bibnamefont
  {Hackenbroich}}, \bibinfo {author} {\bibfnamefont {A.}~\bibnamefont
  {Hudomal}}, \bibinfo {author} {\bibfnamefont {N.}~\bibnamefont {Schuch}},
  \bibinfo {author} {\bibfnamefont {B.~A.}\ \bibnamefont {Bernevig}},\ and\
  \bibinfo {author} {\bibfnamefont {N.}~\bibnamefont {Regnault}},\ }\bibfield
  {title} {\bibinfo {title} {Fractional chiral hinge insulator},\ }\href
  {https://doi.org/10.1103/PhysRevB.103.L161110} {\bibfield  {journal}
  {\bibinfo  {journal} {Phys. Rev. B}\ }\textbf {\bibinfo {volume} {103}},\
  \bibinfo {pages} {L161110} (\bibinfo {year} {2021})}\BibitemShut {NoStop}%
\bibitem [{\citenamefont {Wang}\ \emph {et~al.}(2021)\citenamefont {Wang},
  \citenamefont {Yang}, \citenamefont {Dai},\ and\ \citenamefont
  {Xu}}]{wang2021structural}%
  \BibitemOpen
  \bibfield  {author} {\bibinfo {author} {\bibfnamefont {J.-H.}\ \bibnamefont
  {Wang}}, \bibinfo {author} {\bibfnamefont {Y.-B.}\ \bibnamefont {Yang}},
  \bibinfo {author} {\bibfnamefont {N.}~\bibnamefont {Dai}},\ and\ \bibinfo
  {author} {\bibfnamefont {Y.}~\bibnamefont {Xu}},\ }\bibfield  {title}
  {\bibinfo {title} {Structural-disorder-induced second-order topological
  insulators in three dimensions},\ }\href
  {https://doi.org/10.1103/PhysRevLett.126.206404} {\bibfield  {journal}
  {\bibinfo  {journal} {Phys. Rev. Lett.}\ }\textbf {\bibinfo {volume} {126}},\
  \bibinfo {pages} {206404} (\bibinfo {year} {2021})}\BibitemShut {NoStop}%
\bibitem [{\citenamefont {Fu}\ \emph {et~al.}(2021)\citenamefont {Fu},
  \citenamefont {Hu},\ and\ \citenamefont {Shen}}]{fu2021bulk}%
  \BibitemOpen
  \bibfield  {author} {\bibinfo {author} {\bibfnamefont {B.}~\bibnamefont
  {Fu}}, \bibinfo {author} {\bibfnamefont {Z.-A.}\ \bibnamefont {Hu}},\ and\
  \bibinfo {author} {\bibfnamefont {S.-Q.}\ \bibnamefont {Shen}},\ }\bibfield
  {title} {\bibinfo {title} {Bulk-hinge correspondence and three-dimensional
  quantum anomalous hall effect in second-order topological insulators},\
  }\href {https://doi.org/10.1103/PhysRevResearch.3.033177} {\bibfield
  {journal} {\bibinfo  {journal} {Phys. Rev. Res.}\ }\textbf {\bibinfo {volume}
  {3}},\ \bibinfo {pages} {033177} (\bibinfo {year} {2021})}\BibitemShut
  {NoStop}%
\bibitem [{\citenamefont {Schindler}\ \emph
  {et~al.}(2018{\natexlab{a}})\citenamefont {Schindler}, \citenamefont {Cook},
  \citenamefont {Vergniory}, \citenamefont {Wang}, \citenamefont {Parkin},
  \citenamefont {Bernevig},\ and\ \citenamefont
  {Neupert}}]{schindler2018higher}%
  \BibitemOpen
  \bibfield  {author} {\bibinfo {author} {\bibfnamefont {F.}~\bibnamefont
  {Schindler}}, \bibinfo {author} {\bibfnamefont {A.~M.}\ \bibnamefont {Cook}},
  \bibinfo {author} {\bibfnamefont {M.~G.}\ \bibnamefont {Vergniory}}, \bibinfo
  {author} {\bibfnamefont {Z.}~\bibnamefont {Wang}}, \bibinfo {author}
  {\bibfnamefont {S.~S.}\ \bibnamefont {Parkin}}, \bibinfo {author}
  {\bibfnamefont {B.~A.}\ \bibnamefont {Bernevig}},\ and\ \bibinfo {author}
  {\bibfnamefont {T.}~\bibnamefont {Neupert}},\ }\bibfield  {title} {\bibinfo
  {title} {Higher-order topological insulators},\ }\href
  {https://doi.org/10.1126/sciadv.aat0346} {\bibfield  {journal} {\bibinfo
  {journal} {Sci. Adv.}\ }\textbf {\bibinfo {volume} {4}},\ \bibinfo {pages}
  {eaat0346} (\bibinfo {year} {2018}{\natexlab{a}})}\BibitemShut {NoStop}%
\bibitem [{\citenamefont {Hsieh}\ \emph {et~al.}(2012)\citenamefont {Hsieh},
  \citenamefont {Lin}, \citenamefont {Liu}, \citenamefont {Duan}, \citenamefont
  {Bansil},\ and\ \citenamefont {Fu}}]{hsieh2012topological}%
  \BibitemOpen
  \bibfield  {author} {\bibinfo {author} {\bibfnamefont {T.~H.}\ \bibnamefont
  {Hsieh}}, \bibinfo {author} {\bibfnamefont {H.}~\bibnamefont {Lin}}, \bibinfo
  {author} {\bibfnamefont {J.}~\bibnamefont {Liu}}, \bibinfo {author}
  {\bibfnamefont {W.}~\bibnamefont {Duan}}, \bibinfo {author} {\bibfnamefont
  {A.}~\bibnamefont {Bansil}},\ and\ \bibinfo {author} {\bibfnamefont
  {L.}~\bibnamefont {Fu}},\ }\bibfield  {title} {\bibinfo {title} {Topological
  crystalline insulators in the snte material class},\ }\href
  {https://doi.org/10.1038/ncomms1969} {\bibfield  {journal} {\bibinfo
  {journal} {Nat. Commun.}\ }\textbf {\bibinfo {volume} {3}},\ \bibinfo {pages}
  {982} (\bibinfo {year} {2012})}\BibitemShut {NoStop}%
\bibitem [{\citenamefont {Tanaka}\ \emph {et~al.}(2012)\citenamefont {Tanaka},
  \citenamefont {Ren}, \citenamefont {Sato}, \citenamefont {Nakayama},
  \citenamefont {Souma}, \citenamefont {Takahashi}, \citenamefont {Segawa},\
  and\ \citenamefont {Ando}}]{tanaka2012experimental}%
  \BibitemOpen
  \bibfield  {author} {\bibinfo {author} {\bibfnamefont {Y.}~\bibnamefont
  {Tanaka}}, \bibinfo {author} {\bibfnamefont {Z.}~\bibnamefont {Ren}},
  \bibinfo {author} {\bibfnamefont {T.}~\bibnamefont {Sato}}, \bibinfo {author}
  {\bibfnamefont {K.}~\bibnamefont {Nakayama}}, \bibinfo {author}
  {\bibfnamefont {S.}~\bibnamefont {Souma}}, \bibinfo {author} {\bibfnamefont
  {T.}~\bibnamefont {Takahashi}}, \bibinfo {author} {\bibfnamefont
  {K.}~\bibnamefont {Segawa}},\ and\ \bibinfo {author} {\bibfnamefont
  {Y.}~\bibnamefont {Ando}},\ }\bibfield  {title} {\bibinfo {title}
  {Experimental realization of a topological crystalline insulator in snte},\
  }\href {https://doi.org/10.1038/nphys2442} {\bibfield  {journal} {\bibinfo
  {journal} {Nat. Phys.}\ }\textbf {\bibinfo {volume} {8}},\ \bibinfo {pages}
  {800} (\bibinfo {year} {2012})}\BibitemShut {NoStop}%
\bibitem [{\citenamefont {Benalcazar}\ \emph
  {et~al.}(2017{\natexlab{b}})\citenamefont {Benalcazar}, \citenamefont
  {Bernevig},\ and\ \citenamefont {Hughes}}]{benalcazar2017electric}%
  \BibitemOpen
  \bibfield  {author} {\bibinfo {author} {\bibfnamefont {W.~A.}\ \bibnamefont
  {Benalcazar}}, \bibinfo {author} {\bibfnamefont {B.~A.}\ \bibnamefont
  {Bernevig}},\ and\ \bibinfo {author} {\bibfnamefont {T.~L.}\ \bibnamefont
  {Hughes}},\ }\bibfield  {title} {\bibinfo {title} {Electric multipole
  moments, topological multipole moment pumping, and chiral hinge states in
  crystalline insulators},\ }\href {https://doi.org/10.1103/PhysRevB.96.245115}
  {\bibfield  {journal} {\bibinfo  {journal} {Phys. Rev. B}\ }\textbf {\bibinfo
  {volume} {96}},\ \bibinfo {pages} {245115} (\bibinfo {year}
  {2017}{\natexlab{b}})}\BibitemShut {NoStop}%
\bibitem [{\citenamefont {Sessi}\ \emph {et~al.}(2016)\citenamefont {Sessi},
  \citenamefont {Sante}, \citenamefont {Szczerbakow}, \citenamefont {Glott},
  \citenamefont {Wilfert}, \citenamefont {Schmidt}, \citenamefont {Bathon},
  \citenamefont {Dziawa}, \citenamefont {Greiter}, \citenamefont {Neupert},
  \citenamefont {Sangiovanni}, \citenamefont {Story}, \citenamefont {Thomale},\
  and\ \citenamefont {Bode}}]{Sessi2016RobustSM}%
  \BibitemOpen
  \bibfield  {author} {\bibinfo {author} {\bibfnamefont {P.}~\bibnamefont
  {Sessi}}, \bibinfo {author} {\bibfnamefont {D.~D.}\ \bibnamefont {Sante}},
  \bibinfo {author} {\bibfnamefont {A.}~\bibnamefont {Szczerbakow}}, \bibinfo
  {author} {\bibfnamefont {F.}~\bibnamefont {Glott}}, \bibinfo {author}
  {\bibfnamefont {S.}~\bibnamefont {Wilfert}}, \bibinfo {author} {\bibfnamefont
  {H.}~\bibnamefont {Schmidt}}, \bibinfo {author} {\bibfnamefont
  {T.}~\bibnamefont {Bathon}}, \bibinfo {author} {\bibfnamefont
  {P.}~\bibnamefont {Dziawa}}, \bibinfo {author} {\bibfnamefont
  {M.}~\bibnamefont {Greiter}}, \bibinfo {author} {\bibfnamefont
  {T.}~\bibnamefont {Neupert}}, \bibinfo {author} {\bibfnamefont
  {G.}~\bibnamefont {Sangiovanni}}, \bibinfo {author} {\bibfnamefont
  {T.}~\bibnamefont {Story}}, \bibinfo {author} {\bibfnamefont
  {R.}~\bibnamefont {Thomale}},\ and\ \bibinfo {author} {\bibfnamefont
  {M.}~\bibnamefont {Bode}},\ }\bibfield  {title} {\bibinfo {title} {Robust
  spin-polarized midgap states at step edges of topological crystalline
  insulators},\ }\href {https://doi.org/10.1126/science.aah6233} {\bibfield
  {journal} {\bibinfo  {journal} {Science}\ }\textbf {\bibinfo {volume}
  {354}},\ \bibinfo {pages} {1269 } (\bibinfo {year} {2016})}\BibitemShut
  {NoStop}%
\bibitem [{\citenamefont {Bernevig}\ and\ \citenamefont
  {Zhang}(2006)}]{bernevig2006quantum}%
  \BibitemOpen
  \bibfield  {author} {\bibinfo {author} {\bibfnamefont {B.~A.}\ \bibnamefont
  {Bernevig}}\ and\ \bibinfo {author} {\bibfnamefont {S.-C.}\ \bibnamefont
  {Zhang}},\ }\bibfield  {title} {\bibinfo {title} {Quantum spin hall effect},\
  }\href {https://doi.org/10.1103/PhysRevLett.96.106802} {\bibfield  {journal}
  {\bibinfo  {journal} {Phys. Rev. Lett.}\ }\textbf {\bibinfo {volume} {96}},\
  \bibinfo {pages} {106802} (\bibinfo {year} {2006})}\BibitemShut {NoStop}%
\bibitem [{\citenamefont {Fang}\ and\ \citenamefont {Fu}(2019)}]{fang2019new}%
  \BibitemOpen
  \bibfield  {author} {\bibinfo {author} {\bibfnamefont {C.}~\bibnamefont
  {Fang}}\ and\ \bibinfo {author} {\bibfnamefont {L.}~\bibnamefont {Fu}},\
  }\bibfield  {title} {\bibinfo {title} {New classes of topological crystalline
  insulators having surface rotation anomaly},\ }\href
  {https://doi.org/10.1126/sciadv.aat2374} {\bibfield  {journal} {\bibinfo
  {journal} {Sci. Adv.}\ }\textbf {\bibinfo {volume} {5}},\ \bibinfo {pages}
  {eaat2374} (\bibinfo {year} {2019})}\BibitemShut {NoStop}%
\bibitem [{\citenamefont {Kohda}\ \emph {et~al.}(2019)\citenamefont {Kohda},
  \citenamefont {Okayasu},\ and\ \citenamefont {Nitta}}]{kohda2019spin}%
  \BibitemOpen
  \bibfield  {author} {\bibinfo {author} {\bibfnamefont {M.}~\bibnamefont
  {Kohda}}, \bibinfo {author} {\bibfnamefont {T.}~\bibnamefont {Okayasu}},\
  and\ \bibinfo {author} {\bibfnamefont {J.}~\bibnamefont {Nitta}},\ }\bibfield
   {title} {\bibinfo {title} {Spin-momentum locked spin manipulation in a
  two-dimensional rashba system},\ }\href
  {https://doi.org/10.1038/s41598-018-37967-9} {\bibfield  {journal} {\bibinfo
  {journal} {Sci. Rep.}\ }\textbf {\bibinfo {volume} {9}},\ \bibinfo {pages}
  {1909} (\bibinfo {year} {2019})}\BibitemShut {NoStop}%
\bibitem [{\citenamefont {Yang}\ \emph
  {et~al.}(2020{\natexlab{a}})\citenamefont {Yang}, \citenamefont {Shao},
  \citenamefont {Chen}, \citenamefont {Mao},\ and\ \citenamefont
  {Ma}}]{yang2020spin}%
  \BibitemOpen
  \bibfield  {author} {\bibinfo {author} {\bibfnamefont {Z.-Q.}\ \bibnamefont
  {Yang}}, \bibinfo {author} {\bibfnamefont {Z.-K.}\ \bibnamefont {Shao}},
  \bibinfo {author} {\bibfnamefont {H.-Z.}\ \bibnamefont {Chen}}, \bibinfo
  {author} {\bibfnamefont {X.-R.}\ \bibnamefont {Mao}},\ and\ \bibinfo {author}
  {\bibfnamefont {R.-M.}\ \bibnamefont {Ma}},\ }\bibfield  {title} {\bibinfo
  {title} {Spin-momentum-locked edge mode for topological vortex lasing},\
  }\href {https://doi.org/10.1103/PhysRevLett.125.013903} {\bibfield  {journal}
  {\bibinfo  {journal} {Phys. Rev. Lett.}\ }\textbf {\bibinfo {volume} {125}},\
  \bibinfo {pages} {013903} (\bibinfo {year} {2020}{\natexlab{a}})}\BibitemShut
  {NoStop}%
\bibitem [{\citenamefont {Aggarwal}\ \emph {et~al.}(2021)\citenamefont
  {Aggarwal}, \citenamefont {Zhu}, \citenamefont {Hughes},\ and\ \citenamefont
  {Madhavan}}]{aggarwal2021evidence}%
  \BibitemOpen
  \bibfield  {author} {\bibinfo {author} {\bibfnamefont {L.}~\bibnamefont
  {Aggarwal}}, \bibinfo {author} {\bibfnamefont {P.}~\bibnamefont {Zhu}},
  \bibinfo {author} {\bibfnamefont {T.~L.}\ \bibnamefont {Hughes}},\ and\
  \bibinfo {author} {\bibfnamefont {V.}~\bibnamefont {Madhavan}},\ }\bibfield
  {title} {\bibinfo {title} {Evidence for higher order topology in bi and bi0.
  92sb0. 08},\ }\href {https://doi.org/10.1038/s41467-021-24683-8} {\bibfield
  {journal} {\bibinfo  {journal} {Nat. Commun.}\ }\textbf {\bibinfo {volume}
  {12}},\ \bibinfo {pages} {4420} (\bibinfo {year} {2021})}\BibitemShut
  {NoStop}%
\bibitem [{\citenamefont {Schindler}\ \emph
  {et~al.}(2018{\natexlab{b}})\citenamefont {Schindler}, \citenamefont {Wang},
  \citenamefont {Vergniory}, \citenamefont {Cook}, \citenamefont {Murani},
  \citenamefont {Sengupta}, \citenamefont {Kasumov}, \citenamefont {Deblock},
  \citenamefont {Jeon}, \citenamefont {Drozdov} \emph
  {et~al.}}]{schindler2018bismuth}%
  \BibitemOpen
  \bibfield  {author} {\bibinfo {author} {\bibfnamefont {F.}~\bibnamefont
  {Schindler}}, \bibinfo {author} {\bibfnamefont {Z.}~\bibnamefont {Wang}},
  \bibinfo {author} {\bibfnamefont {M.~G.}\ \bibnamefont {Vergniory}}, \bibinfo
  {author} {\bibfnamefont {A.~M.}\ \bibnamefont {Cook}}, \bibinfo {author}
  {\bibfnamefont {A.}~\bibnamefont {Murani}}, \bibinfo {author} {\bibfnamefont
  {S.}~\bibnamefont {Sengupta}}, \bibinfo {author} {\bibfnamefont {A.~Y.}\
  \bibnamefont {Kasumov}}, \bibinfo {author} {\bibfnamefont {R.}~\bibnamefont
  {Deblock}}, \bibinfo {author} {\bibfnamefont {S.}~\bibnamefont {Jeon}},
  \bibinfo {author} {\bibfnamefont {I.}~\bibnamefont {Drozdov}}, \emph
  {et~al.},\ }\bibfield  {title} {\bibinfo {title} {Higher-order topology in
  bismuth},\ }\href {https://doi.org/10.1038/s41567-018-0224-7} {\bibfield
  {journal} {\bibinfo  {journal} {Nat. Phys.}\ }\textbf {\bibinfo {volume}
  {14}},\ \bibinfo {pages} {918} (\bibinfo {year}
  {2018}{\natexlab{b}})}\BibitemShut {NoStop}%
\bibitem [{\citenamefont {Hsu}\ \emph {et~al.}(2019)\citenamefont {Hsu},
  \citenamefont {Zhou}, \citenamefont {Ma}, \citenamefont {Gedik},
  \citenamefont {Bansil}, \citenamefont {Pereira}, \citenamefont {Lin},
  \citenamefont {Fu}, \citenamefont {Xu},\ and\ \citenamefont
  {Chang}}]{hsu2019purely}%
  \BibitemOpen
  \bibfield  {author} {\bibinfo {author} {\bibfnamefont {C.-H.}\ \bibnamefont
  {Hsu}}, \bibinfo {author} {\bibfnamefont {X.}~\bibnamefont {Zhou}}, \bibinfo
  {author} {\bibfnamefont {Q.}~\bibnamefont {Ma}}, \bibinfo {author}
  {\bibfnamefont {N.}~\bibnamefont {Gedik}}, \bibinfo {author} {\bibfnamefont
  {A.}~\bibnamefont {Bansil}}, \bibinfo {author} {\bibfnamefont {V.~M.}\
  \bibnamefont {Pereira}}, \bibinfo {author} {\bibfnamefont {H.}~\bibnamefont
  {Lin}}, \bibinfo {author} {\bibfnamefont {L.}~\bibnamefont {Fu}}, \bibinfo
  {author} {\bibfnamefont {S.-Y.}\ \bibnamefont {Xu}},\ and\ \bibinfo {author}
  {\bibfnamefont {T.-R.}\ \bibnamefont {Chang}},\ }\bibfield  {title} {\bibinfo
  {title} {Purely rotational symmetry-protected topological crystalline
  insulator-bi4br4},\ }\href {https://doi.org/10.1088/2053-1583/ab1607}
  {\bibfield  {journal} {\bibinfo  {journal} {2D Mater.}\ }\textbf {\bibinfo
  {volume} {6}},\ \bibinfo {pages} {031004} (\bibinfo {year}
  {2019})}\BibitemShut {NoStop}%
\bibitem [{\citenamefont {Shumiya}\ \emph {et~al.}(2022)\citenamefont
  {Shumiya}, \citenamefont {Hossain}, \citenamefont {Yin}, \citenamefont
  {Wang}, \citenamefont {Litskevich}, \citenamefont {Yoon}, \citenamefont {Li},
  \citenamefont {Yang}, \citenamefont {Jiang}, \citenamefont {Cheng} \emph
  {et~al.}}]{shumiya2022evidence}%
  \BibitemOpen
  \bibfield  {author} {\bibinfo {author} {\bibfnamefont {N.}~\bibnamefont
  {Shumiya}}, \bibinfo {author} {\bibfnamefont {M.~S.}\ \bibnamefont
  {Hossain}}, \bibinfo {author} {\bibfnamefont {J.-X.}\ \bibnamefont {Yin}},
  \bibinfo {author} {\bibfnamefont {Z.}~\bibnamefont {Wang}}, \bibinfo {author}
  {\bibfnamefont {M.}~\bibnamefont {Litskevich}}, \bibinfo {author}
  {\bibfnamefont {C.}~\bibnamefont {Yoon}}, \bibinfo {author} {\bibfnamefont
  {Y.}~\bibnamefont {Li}}, \bibinfo {author} {\bibfnamefont {Y.}~\bibnamefont
  {Yang}}, \bibinfo {author} {\bibfnamefont {Y.-X.}\ \bibnamefont {Jiang}},
  \bibinfo {author} {\bibfnamefont {G.}~\bibnamefont {Cheng}}, \emph {et~al.},\
  }\bibfield  {title} {\bibinfo {title} {Evidence of a room-temperature quantum
  spin hall edge state in a higher-order topological insulator},\ }\href
  {https://doi.org/10.1038/s41563-022-01304-3} {\bibfield  {journal} {\bibinfo
  {journal} {Nat. Mater.}\ }\textbf {\bibinfo {volume} {21}},\ \bibinfo {pages}
  {1111} (\bibinfo {year} {2022})}\BibitemShut {NoStop}%
\bibitem [{\citenamefont {Wang}\ \emph {et~al.}(2019)\citenamefont {Wang},
  \citenamefont {Wieder}, \citenamefont {Li}, \citenamefont {Yan},\ and\
  \citenamefont {Bernevig}}]{wang2019higher}%
  \BibitemOpen
  \bibfield  {author} {\bibinfo {author} {\bibfnamefont {Z.}~\bibnamefont
  {Wang}}, \bibinfo {author} {\bibfnamefont {B.~J.}\ \bibnamefont {Wieder}},
  \bibinfo {author} {\bibfnamefont {J.}~\bibnamefont {Li}}, \bibinfo {author}
  {\bibfnamefont {B.}~\bibnamefont {Yan}},\ and\ \bibinfo {author}
  {\bibfnamefont {B.~A.}\ \bibnamefont {Bernevig}},\ }\bibfield  {title}
  {\bibinfo {title} {Higher-order topology, monopole nodal lines, and the
  origin of large fermi arcs in transition metal dichalcogenides
  $x{\mathrm{te}}_{2}$ ($x=\mathrm{Mo},\mathrm{W}$)},\ }\href
  {https://doi.org/10.1103/PhysRevLett.123.186401} {\bibfield  {journal}
  {\bibinfo  {journal} {Phys. Rev. Lett.}\ }\textbf {\bibinfo {volume} {123}},\
  \bibinfo {pages} {186401} (\bibinfo {year} {2019})}\BibitemShut {NoStop}%
\bibitem [{\citenamefont {Noguchi}\ \emph {et~al.}(2021)\citenamefont
  {Noguchi}, \citenamefont {Kobayashi}, \citenamefont {Jiang}, \citenamefont
  {Kuroda}, \citenamefont {Takahashi}, \citenamefont {Xu}, \citenamefont {Lee},
  \citenamefont {Hirayama}, \citenamefont {Ochi}, \citenamefont {Shirasawa}
  \emph {et~al.}}]{noguchi2021evidence}%
  \BibitemOpen
  \bibfield  {author} {\bibinfo {author} {\bibfnamefont {R.}~\bibnamefont
  {Noguchi}}, \bibinfo {author} {\bibfnamefont {M.}~\bibnamefont {Kobayashi}},
  \bibinfo {author} {\bibfnamefont {Z.}~\bibnamefont {Jiang}}, \bibinfo
  {author} {\bibfnamefont {K.}~\bibnamefont {Kuroda}}, \bibinfo {author}
  {\bibfnamefont {T.}~\bibnamefont {Takahashi}}, \bibinfo {author}
  {\bibfnamefont {Z.}~\bibnamefont {Xu}}, \bibinfo {author} {\bibfnamefont
  {D.}~\bibnamefont {Lee}}, \bibinfo {author} {\bibfnamefont {M.}~\bibnamefont
  {Hirayama}}, \bibinfo {author} {\bibfnamefont {M.}~\bibnamefont {Ochi}},
  \bibinfo {author} {\bibfnamefont {T.}~\bibnamefont {Shirasawa}}, \emph
  {et~al.},\ }\bibfield  {title} {\bibinfo {title} {Evidence for a higher-order
  topological insulator in a three-dimensional material built from van der
  waals stacking of bismuth-halide chains},\ }\href
  {https://doi.org/10.1038/s41563-020-00871-7} {\bibfield  {journal} {\bibinfo
  {journal} {Nat. Mater.}\ }\textbf {\bibinfo {volume} {20}},\ \bibinfo {pages}
  {473} (\bibinfo {year} {2021})}\BibitemShut {NoStop}%
\bibitem [{\citenamefont {Laubscher}\ \emph
  {et~al.}(2023{\natexlab{a}})\citenamefont {Laubscher}, \citenamefont
  {Keizer},\ and\ \citenamefont {Klinovaja}}]{laubscher2023fractional}%
  \BibitemOpen
  \bibfield  {author} {\bibinfo {author} {\bibfnamefont {K.}~\bibnamefont
  {Laubscher}}, \bibinfo {author} {\bibfnamefont {P.}~\bibnamefont {Keizer}},\
  and\ \bibinfo {author} {\bibfnamefont {J.}~\bibnamefont {Klinovaja}},\
  }\bibfield  {title} {\bibinfo {title} {Fractional second-order topological
  insulator from a three-dimensional coupled-wires construction},\ }\href
  {https://doi.org/10.1103/PhysRevB.107.045409} {\bibfield  {journal} {\bibinfo
   {journal} {Phys. Rev. B}\ }\textbf {\bibinfo {volume} {107}},\ \bibinfo
  {pages} {045409} (\bibinfo {year} {2023}{\natexlab{a}})}\BibitemShut
  {NoStop}%
\bibitem [{\citenamefont {Yang}\ \emph
  {et~al.}(2020{\natexlab{b}})\citenamefont {Yang}, \citenamefont {Li},
  \citenamefont {Peng}, \citenamefont {Zou},\ and\ \citenamefont
  {Cheng}}]{yang2020helical}%
  \BibitemOpen
  \bibfield  {author} {\bibinfo {author} {\bibfnamefont {Z.-Z.}\ \bibnamefont
  {Yang}}, \bibinfo {author} {\bibfnamefont {X.}~\bibnamefont {Li}}, \bibinfo
  {author} {\bibfnamefont {Y.-Y.}\ \bibnamefont {Peng}}, \bibinfo {author}
  {\bibfnamefont {X.-Y.}\ \bibnamefont {Zou}},\ and\ \bibinfo {author}
  {\bibfnamefont {J.-C.}\ \bibnamefont {Cheng}},\ }\bibfield  {title} {\bibinfo
  {title} {Helical higher-order topological states in an acoustic crystalline
  insulator},\ }\href {https://doi.org/10.1103/PhysRevLett.125.255502}
  {\bibfield  {journal} {\bibinfo  {journal} {Phys. Rev. Lett.}\ }\textbf
  {\bibinfo {volume} {125}},\ \bibinfo {pages} {255502} (\bibinfo {year}
  {2020}{\natexlab{b}})}\BibitemShut {NoStop}%
\bibitem [{\citenamefont {Craig}\ \emph {et~al.}(2004)\citenamefont {Craig},
  \citenamefont {Taylor}, \citenamefont {Lester}, \citenamefont {Marcus},
  \citenamefont {Hanson},\ and\ \citenamefont {Gossard}}]{craig2004tunable}%
  \BibitemOpen
  \bibfield  {author} {\bibinfo {author} {\bibfnamefont {N.}~\bibnamefont
  {Craig}}, \bibinfo {author} {\bibfnamefont {J.}~\bibnamefont {Taylor}},
  \bibinfo {author} {\bibfnamefont {E.}~\bibnamefont {Lester}}, \bibinfo
  {author} {\bibfnamefont {C.}~\bibnamefont {Marcus}}, \bibinfo {author}
  {\bibfnamefont {M.}~\bibnamefont {Hanson}},\ and\ \bibinfo {author}
  {\bibfnamefont {A.}~\bibnamefont {Gossard}},\ }\bibfield  {title} {\bibinfo
  {title} {Tunable nonlocal spin control in a coupled-quantum dot system},\
  }\href {https://doi.org/10.1126/science.1095452} {\bibfield  {journal}
  {\bibinfo  {journal} {Science}\ }\textbf {\bibinfo {volume} {304}},\ \bibinfo
  {pages} {565} (\bibinfo {year} {2004})}\BibitemShut {NoStop}%
\bibitem [{\citenamefont {Usaj}\ \emph {et~al.}(2005)\citenamefont {Usaj},
  \citenamefont {Lustemberg},\ and\ \citenamefont {Balseiro}}]{usaj2005tuning}%
  \BibitemOpen
  \bibfield  {author} {\bibinfo {author} {\bibfnamefont {G.}~\bibnamefont
  {Usaj}}, \bibinfo {author} {\bibfnamefont {P.}~\bibnamefont {Lustemberg}},\
  and\ \bibinfo {author} {\bibfnamefont {C.~A.}\ \bibnamefont {Balseiro}},\
  }\bibfield  {title} {\bibinfo {title} {Tuning the nonlocal spin-spin
  interaction between quantum dots with a magnetic field},\ }\href
  {https://doi.org/10.1103/PhysRevLett.94.036803} {\bibfield  {journal}
  {\bibinfo  {journal} {Phys. Rev. Lett.}\ }\textbf {\bibinfo {volume} {94}},\
  \bibinfo {pages} {036803} (\bibinfo {year} {2005})}\BibitemShut {NoStop}%
\bibitem [{\citenamefont {Dugaev}\ \emph {et~al.}(2006)\citenamefont {Dugaev},
  \citenamefont {Litvinov},\ and\ \citenamefont {Barnas}}]{dugaev2006exchange}%
  \BibitemOpen
  \bibfield  {author} {\bibinfo {author} {\bibfnamefont {V.~K.}\ \bibnamefont
  {Dugaev}}, \bibinfo {author} {\bibfnamefont {V.~I.}\ \bibnamefont
  {Litvinov}},\ and\ \bibinfo {author} {\bibfnamefont {J.}~\bibnamefont
  {Barnas}},\ }\bibfield  {title} {\bibinfo {title} {Exchange interaction of
  magnetic impurities in graphene},\ }\href
  {https://doi.org/10.1103/PhysRevB.74.224438} {\bibfield  {journal} {\bibinfo
  {journal} {Phys. Rev. B}\ }\textbf {\bibinfo {volume} {74}},\ \bibinfo
  {pages} {224438} (\bibinfo {year} {2006})}\BibitemShut {NoStop}%
\bibitem [{\citenamefont {Brey}\ \emph {et~al.}(2007)\citenamefont {Brey},
  \citenamefont {Fertig},\ and\ \citenamefont {Das~Sarma}}]{brey2007diluted}%
  \BibitemOpen
  \bibfield  {author} {\bibinfo {author} {\bibfnamefont {L.}~\bibnamefont
  {Brey}}, \bibinfo {author} {\bibfnamefont {H.~A.}\ \bibnamefont {Fertig}},\
  and\ \bibinfo {author} {\bibfnamefont {S.}~\bibnamefont {Das~Sarma}},\
  }\bibfield  {title} {\bibinfo {title} {Diluted graphene antiferromagnet},\
  }\href {https://doi.org/10.1103/PhysRevLett.99.116802} {\bibfield  {journal}
  {\bibinfo  {journal} {Phys. Rev. Lett.}\ }\textbf {\bibinfo {volume} {99}},\
  \bibinfo {pages} {116802} (\bibinfo {year} {2007})}\BibitemShut {NoStop}%
\bibitem [{\citenamefont {Liu}\ \emph {et~al.}(2009)\citenamefont {Liu},
  \citenamefont {Liu}, \citenamefont {Xu}, \citenamefont {Qi},\ and\
  \citenamefont {Zhang}}]{liu2009magnetic}%
  \BibitemOpen
  \bibfield  {author} {\bibinfo {author} {\bibfnamefont {Q.}~\bibnamefont
  {Liu}}, \bibinfo {author} {\bibfnamefont {C.-X.}\ \bibnamefont {Liu}},
  \bibinfo {author} {\bibfnamefont {C.}~\bibnamefont {Xu}}, \bibinfo {author}
  {\bibfnamefont {X.-L.}\ \bibnamefont {Qi}},\ and\ \bibinfo {author}
  {\bibfnamefont {S.-C.}\ \bibnamefont {Zhang}},\ }\bibfield  {title} {\bibinfo
  {title} {Magnetic impurities on the surface of a topological insulator},\
  }\href {https://doi.org/10.1103/PhysRevLett.102.156603} {\bibfield  {journal}
  {\bibinfo  {journal} {Phys. Rev. Lett.}\ }\textbf {\bibinfo {volume} {102}},\
  \bibinfo {pages} {156603} (\bibinfo {year} {2009})}\BibitemShut {NoStop}%
\bibitem [{\citenamefont {Zhu}\ \emph {et~al.}(2011)\citenamefont {Zhu},
  \citenamefont {Yao}, \citenamefont {Zhang},\ and\ \citenamefont
  {Chang}}]{zhu2011electrically}%
  \BibitemOpen
  \bibfield  {author} {\bibinfo {author} {\bibfnamefont {J.-J.}\ \bibnamefont
  {Zhu}}, \bibinfo {author} {\bibfnamefont {D.-X.}\ \bibnamefont {Yao}},
  \bibinfo {author} {\bibfnamefont {S.-C.}\ \bibnamefont {Zhang}},\ and\
  \bibinfo {author} {\bibfnamefont {K.}~\bibnamefont {Chang}},\ }\bibfield
  {title} {\bibinfo {title} {Electrically controllable surface magnetism on the
  surface of topological insulators},\ }\href
  {https://doi.org/10.1103/PhysRevLett.106.097201} {\bibfield  {journal}
  {\bibinfo  {journal} {Phys. Rev. Lett.}\ }\textbf {\bibinfo {volume} {106}},\
  \bibinfo {pages} {097201} (\bibinfo {year} {2011})}\BibitemShut {NoStop}%
\bibitem [{\citenamefont {Chang}\ \emph {et~al.}(2015)\citenamefont {Chang},
  \citenamefont {Zhou}, \citenamefont {Wang}, \citenamefont {Shan},\ and\
  \citenamefont {Xiao}}]{chang2015rkky}%
  \BibitemOpen
  \bibfield  {author} {\bibinfo {author} {\bibfnamefont {H.-R.}\ \bibnamefont
  {Chang}}, \bibinfo {author} {\bibfnamefont {J.}~\bibnamefont {Zhou}},
  \bibinfo {author} {\bibfnamefont {S.-X.}\ \bibnamefont {Wang}}, \bibinfo
  {author} {\bibfnamefont {W.-Y.}\ \bibnamefont {Shan}},\ and\ \bibinfo
  {author} {\bibfnamefont {D.}~\bibnamefont {Xiao}},\ }\bibfield  {title}
  {\bibinfo {title} {Rkky interaction of magnetic impurities in dirac and weyl
  semimetals},\ }\href {https://doi.org/10.1103/PhysRevB.92.241103} {\bibfield
  {journal} {\bibinfo  {journal} {Phys. Rev. B}\ }\textbf {\bibinfo {volume}
  {92}},\ \bibinfo {pages} {241103} (\bibinfo {year} {2015})}\BibitemShut
  {NoStop}%
\bibitem [{\citenamefont {Hosseini}\ and\ \citenamefont
  {Askari}(2015)}]{hosseini2015ruderman}%
  \BibitemOpen
  \bibfield  {author} {\bibinfo {author} {\bibfnamefont {M.~V.}\ \bibnamefont
  {Hosseini}}\ and\ \bibinfo {author} {\bibfnamefont {M.}~\bibnamefont
  {Askari}},\ }\bibfield  {title} {\bibinfo {title}
  {Ruderman-kittel-kasuya-yosida interaction in weyl semimetals},\ }\href
  {https://doi.org/10.1103/PhysRevB.92.224435} {\bibfield  {journal} {\bibinfo
  {journal} {Phys. Rev. B}\ }\textbf {\bibinfo {volume} {92}},\ \bibinfo
  {pages} {224435} (\bibinfo {year} {2015})}\BibitemShut {NoStop}%
\bibitem [{\citenamefont {Yarmohammadi}\ and\ \citenamefont
  {Cheraghchi}(2020)}]{yarmohammadi2020effective}%
  \BibitemOpen
  \bibfield  {author} {\bibinfo {author} {\bibfnamefont {M.}~\bibnamefont
  {Yarmohammadi}}\ and\ \bibinfo {author} {\bibfnamefont {H.}~\bibnamefont
  {Cheraghchi}},\ }\bibfield  {title} {\bibinfo {title} {Effective low-energy
  rkky interaction in doped topological crystalline insulators},\ }\href
  {https://doi.org/10.1103/PhysRevB.102.075411} {\bibfield  {journal} {\bibinfo
   {journal} {Phys. Rev. B}\ }\textbf {\bibinfo {volume} {102}},\ \bibinfo
  {pages} {075411} (\bibinfo {year} {2020})}\BibitemShut {NoStop}%
\bibitem [{\citenamefont {Cheraghchi}\ and\ \citenamefont
  {Yarmohammadi}(2021)}]{cheraghchi2021anisotropic}%
  \BibitemOpen
  \bibfield  {author} {\bibinfo {author} {\bibfnamefont {H.}~\bibnamefont
  {Cheraghchi}}\ and\ \bibinfo {author} {\bibfnamefont {M.}~\bibnamefont
  {Yarmohammadi}},\ }\bibfield  {title} {\bibinfo {title} {Anisotropic
  ferroelectric distortion effects on the rkky interaction in topological
  crystalline insulators},\ }\href {https://doi.org/10.1038/s41598-021-84398-0}
  {\bibfield  {journal} {\bibinfo  {journal} {Sci. Rep.}\ }\textbf {\bibinfo
  {volume} {11}},\ \bibinfo {pages} {5273} (\bibinfo {year}
  {2021})}\BibitemShut {NoStop}%
\bibitem [{\citenamefont {Yarmohammadi}\ \emph {et~al.}(2023)\citenamefont
  {Yarmohammadi}, \citenamefont {Bukov},\ and\ \citenamefont
  {Kolodrubetz}}]{yarmohammadi2023noncollinear}%
  \BibitemOpen
  \bibfield  {author} {\bibinfo {author} {\bibfnamefont {M.}~\bibnamefont
  {Yarmohammadi}}, \bibinfo {author} {\bibfnamefont {M.}~\bibnamefont
  {Bukov}},\ and\ \bibinfo {author} {\bibfnamefont {M.~H.}\ \bibnamefont
  {Kolodrubetz}},\ }\bibfield  {title} {\bibinfo {title} {Noncollinear twisted
  rkky interaction on the optically driven snte(001) surface},\ }\href
  {https://doi.org/10.1103/PhysRevB.107.054439} {\bibfield  {journal} {\bibinfo
   {journal} {Phys. Rev. B}\ }\textbf {\bibinfo {volume} {107}},\ \bibinfo
  {pages} {054439} (\bibinfo {year} {2023})}\BibitemShut {NoStop}%
\bibitem [{\citenamefont {Hickel}\ and\ \citenamefont
  {Nolting}(2004)}]{hickel2004proper}%
  \BibitemOpen
  \bibfield  {author} {\bibinfo {author} {\bibfnamefont {T.}~\bibnamefont
  {Hickel}}\ and\ \bibinfo {author} {\bibfnamefont {W.}~\bibnamefont
  {Nolting}},\ }\bibfield  {title} {\bibinfo {title} {Proper weak-coupling
  approach to the periodic $s\ensuremath{-}d(f)$ exchange model},\ }\href
  {https://doi.org/10.1103/PhysRevB.69.085110} {\bibfield  {journal} {\bibinfo
  {journal} {Phys. Rev. B}\ }\textbf {\bibinfo {volume} {69}},\ \bibinfo
  {pages} {085110} (\bibinfo {year} {2004})}\BibitemShut {NoStop}%
\bibitem [{\citenamefont {Nolting}\ \emph {et~al.}(2001)\citenamefont
  {Nolting}, \citenamefont {Reddy}, \citenamefont {Ramakanth},\ and\
  \citenamefont {Meyer}}]{nolting2001low}%
  \BibitemOpen
  \bibfield  {author} {\bibinfo {author} {\bibfnamefont {W.}~\bibnamefont
  {Nolting}}, \bibinfo {author} {\bibfnamefont {G.~G.}\ \bibnamefont {Reddy}},
  \bibinfo {author} {\bibfnamefont {A.}~\bibnamefont {Ramakanth}},\ and\
  \bibinfo {author} {\bibfnamefont {D.}~\bibnamefont {Meyer}},\ }\bibfield
  {title} {\bibinfo {title} {Low-density approach to the kondo-lattice model},\
  }\href {https://doi.org/10.1103/PhysRevB.64.155109} {\bibfield  {journal}
  {\bibinfo  {journal} {Phys. Rev. B}\ }\textbf {\bibinfo {volume} {64}},\
  \bibinfo {pages} {155109} (\bibinfo {year} {2001})}\BibitemShut {NoStop}%
\bibitem [{\citenamefont {Schindler}\ \emph {et~al.}(2022)\citenamefont
  {Schindler}, \citenamefont {Tsirkin}, \citenamefont {Neupert}, \citenamefont
  {Andrei~Bernevig},\ and\ \citenamefont {Wieder}}]{schindler2022topological}%
  \BibitemOpen
  \bibfield  {author} {\bibinfo {author} {\bibfnamefont {F.}~\bibnamefont
  {Schindler}}, \bibinfo {author} {\bibfnamefont {S.~S.}\ \bibnamefont
  {Tsirkin}}, \bibinfo {author} {\bibfnamefont {T.}~\bibnamefont {Neupert}},
  \bibinfo {author} {\bibfnamefont {B.}~\bibnamefont {Andrei~Bernevig}},\ and\
  \bibinfo {author} {\bibfnamefont {B.~J.}\ \bibnamefont {Wieder}},\ }\bibfield
   {title} {\bibinfo {title} {Topological zero-dimensional defect and flux
  states in three-dimensional insulators},\ }\href
  {https://doi.org/10.1038/s41467-022-33471-x} {\bibfield  {journal} {\bibinfo
  {journal} {Nat. Commun.}\ }\textbf {\bibinfo {volume} {13}},\ \bibinfo
  {pages} {5791} (\bibinfo {year} {2022})}\BibitemShut {NoStop}%
\bibitem [{\citenamefont {Laubscher}\ \emph
  {et~al.}(2023{\natexlab{b}})\citenamefont {Laubscher}, \citenamefont
  {Miserev}, \citenamefont {Kaladzhyan}, \citenamefont {Loss},\ and\
  \citenamefont {Klinovaja}}]{laubscher2023rkky}%
  \BibitemOpen
  \bibfield  {author} {\bibinfo {author} {\bibfnamefont {K.}~\bibnamefont
  {Laubscher}}, \bibinfo {author} {\bibfnamefont {D.}~\bibnamefont {Miserev}},
  \bibinfo {author} {\bibfnamefont {V.}~\bibnamefont {Kaladzhyan}}, \bibinfo
  {author} {\bibfnamefont {D.}~\bibnamefont {Loss}},\ and\ \bibinfo {author}
  {\bibfnamefont {J.}~\bibnamefont {Klinovaja}},\ }\bibfield  {title} {\bibinfo
  {title} {Rkky interaction at helical edges of topological superconductors},\
  }\href {https://doi.org/10.1103/PhysRevB.107.115421} {\bibfield  {journal}
  {\bibinfo  {journal} {Phys. Rev. B}\ }\textbf {\bibinfo {volume} {107}},\
  \bibinfo {pages} {115421} (\bibinfo {year} {2023}{\natexlab{b}})}\BibitemShut
  {NoStop}%
\bibitem [{\citenamefont {Schmidt}\ \emph {et~al.}(2011)\citenamefont
  {Schmidt}, \citenamefont {Miwa},\ and\ \citenamefont
  {Fazzio}}]{PhysRevB.84.245418}%
  \BibitemOpen
  \bibfield  {author} {\bibinfo {author} {\bibfnamefont {T.~M.}\ \bibnamefont
  {Schmidt}}, \bibinfo {author} {\bibfnamefont {R.~H.}\ \bibnamefont {Miwa}},\
  and\ \bibinfo {author} {\bibfnamefont {A.}~\bibnamefont {Fazzio}},\
  }\bibfield  {title} {\bibinfo {title} {Spin texture and magnetic anisotropy
  of co impurities in bi${}_{2}$se${}_{3}$ topological insulators},\ }\href
  {https://doi.org/10.1103/PhysRevB.84.245418} {\bibfield  {journal} {\bibinfo
  {journal} {Phys. Rev. B}\ }\textbf {\bibinfo {volume} {84}},\ \bibinfo
  {pages} {245418} (\bibinfo {year} {2011})}\BibitemShut {NoStop}%
\bibitem [{\citenamefont {Henk}\ \emph {et~al.}(2012)\citenamefont {Henk},
  \citenamefont {Ernst}, \citenamefont {Eremeev}, \citenamefont {Chulkov},
  \citenamefont {Maznichenko},\ and\ \citenamefont
  {Mertig}}]{PhysRevLett.108.206801}%
  \BibitemOpen
  \bibfield  {author} {\bibinfo {author} {\bibfnamefont {J.}~\bibnamefont
  {Henk}}, \bibinfo {author} {\bibfnamefont {A.}~\bibnamefont {Ernst}},
  \bibinfo {author} {\bibfnamefont {S.~V.}\ \bibnamefont {Eremeev}}, \bibinfo
  {author} {\bibfnamefont {E.~V.}\ \bibnamefont {Chulkov}}, \bibinfo {author}
  {\bibfnamefont {I.~V.}\ \bibnamefont {Maznichenko}},\ and\ \bibinfo {author}
  {\bibfnamefont {I.}~\bibnamefont {Mertig}},\ }\bibfield  {title} {\bibinfo
  {title} {Complex spin texture in the pure and mn-doped topological insulator
  ${\mathrm{bi}}_{2}{\mathrm{te}}_{3}$},\ }\href
  {https://doi.org/10.1103/PhysRevLett.108.206801} {\bibfield  {journal}
  {\bibinfo  {journal} {Phys. Rev. Lett.}\ }\textbf {\bibinfo {volume} {108}},\
  \bibinfo {pages} {206801} (\bibinfo {year} {2012})}\BibitemShut {NoStop}%
\bibitem [{\citenamefont {Zhang}\ \emph {et~al.}(2012)\citenamefont {Zhang},
  \citenamefont {Zhu}, \citenamefont {Zhang}, \citenamefont {Xiao},\ and\
  \citenamefont {Yao}}]{PhysRevLett.109.266405}%
  \BibitemOpen
  \bibfield  {author} {\bibinfo {author} {\bibfnamefont {J.-M.}\ \bibnamefont
  {Zhang}}, \bibinfo {author} {\bibfnamefont {W.}~\bibnamefont {Zhu}}, \bibinfo
  {author} {\bibfnamefont {Y.}~\bibnamefont {Zhang}}, \bibinfo {author}
  {\bibfnamefont {D.}~\bibnamefont {Xiao}},\ and\ \bibinfo {author}
  {\bibfnamefont {Y.}~\bibnamefont {Yao}},\ }\bibfield  {title} {\bibinfo
  {title} {Tailoring magnetic doping in the topological insulator
  ${\mathrm{bi}}_{2}{\mathrm{se}}_{3}$},\ }\href
  {https://doi.org/10.1103/PhysRevLett.109.266405} {\bibfield  {journal}
  {\bibinfo  {journal} {Phys. Rev. Lett.}\ }\textbf {\bibinfo {volume} {109}},\
  \bibinfo {pages} {266405} (\bibinfo {year} {2012})}\BibitemShut {NoStop}%
\bibitem [{\citenamefont {Sun}\ \emph {et~al.}(2014)\citenamefont {Sun},
  \citenamefont {Chen},\ and\ \citenamefont {Lin}}]{PhysRevB.89.115101}%
  \BibitemOpen
  \bibfield  {author} {\bibinfo {author} {\bibfnamefont {J.}~\bibnamefont
  {Sun}}, \bibinfo {author} {\bibfnamefont {L.}~\bibnamefont {Chen}},\ and\
  \bibinfo {author} {\bibfnamefont {H.-Q.}\ \bibnamefont {Lin}},\ }\bibfield
  {title} {\bibinfo {title} {Spin-spin interaction in the bulk of topological
  insulators},\ }\href {https://doi.org/10.1103/PhysRevB.89.115101} {\bibfield
  {journal} {\bibinfo  {journal} {Phys. Rev. B}\ }\textbf {\bibinfo {volume}
  {89}},\ \bibinfo {pages} {115101} (\bibinfo {year} {2014})}\BibitemShut
  {NoStop}%
\bibitem [{\citenamefont {Duan}\ \emph {et~al.}(2022)\citenamefont {Duan},
  \citenamefont {Wu}, \citenamefont {Yang}, \citenamefont {Zheng},
  \citenamefont {Zhu}, \citenamefont {Deng}, \citenamefont {Yang},\ and\
  \citenamefont {Wang}}]{duan2022prolonged}%
  \BibitemOpen
  \bibfield  {author} {\bibinfo {author} {\bibfnamefont {H.-J.}\ \bibnamefont
  {Duan}}, \bibinfo {author} {\bibfnamefont {Y.-J.}\ \bibnamefont {Wu}},
  \bibinfo {author} {\bibfnamefont {Y.-Y.}\ \bibnamefont {Yang}}, \bibinfo
  {author} {\bibfnamefont {S.-H.}\ \bibnamefont {Zheng}}, \bibinfo {author}
  {\bibfnamefont {C.-Y.}\ \bibnamefont {Zhu}}, \bibinfo {author} {\bibfnamefont
  {M.-X.}\ \bibnamefont {Deng}}, \bibinfo {author} {\bibfnamefont
  {M.}~\bibnamefont {Yang}},\ and\ \bibinfo {author} {\bibfnamefont {R.-Q.}\
  \bibnamefont {Wang}},\ }\bibfield  {title} {\bibinfo {title} {The prolonged
  decay of rkky interactions by interplay of relativistic and non-relativistic
  electrons in semi-dirac semimetals},\ }\href
  {https://doi.org/10.1088/1367-2630/ac5842} {\bibfield  {journal} {\bibinfo
  {journal} {New Journal of Physics}\ }\textbf {\bibinfo {volume} {24}},\
  \bibinfo {pages} {033029} (\bibinfo {year} {2022})}\BibitemShut {NoStop}%
\bibitem [{\citenamefont {Mohammadi}\ and\ \citenamefont
  {Moghaddam}(2020)}]{PhysRevB.101.075421}%
  \BibitemOpen
  \bibfield  {author} {\bibinfo {author} {\bibfnamefont {R.~G.}\ \bibnamefont
  {Mohammadi}}\ and\ \bibinfo {author} {\bibfnamefont {A.~G.}\ \bibnamefont
  {Moghaddam}},\ }\bibfield  {title} {\bibinfo {title} {Anisotropic rkky
  interactions mediated by $j=\frac{3}{2}$ quasiparticles in half-heusler
  topological semimetals},\ }\href
  {https://doi.org/10.1103/PhysRevB.101.075421} {\bibfield  {journal} {\bibinfo
   {journal} {Phys. Rev. B}\ }\textbf {\bibinfo {volume} {101}},\ \bibinfo
  {pages} {075421} (\bibinfo {year} {2020})}\BibitemShut {NoStop}%
\bibitem [{\citenamefont {Mross}\ and\ \citenamefont
  {Johannesson}(2009)}]{PhysRevB.80.155302}%
  \BibitemOpen
  \bibfield  {author} {\bibinfo {author} {\bibfnamefont {D.~F.}\ \bibnamefont
  {Mross}}\ and\ \bibinfo {author} {\bibfnamefont {H.}~\bibnamefont
  {Johannesson}},\ }\bibfield  {title} {\bibinfo {title} {Two-impurity kondo
  model with spin-orbit interactions},\ }\href
  {https://doi.org/10.1103/PhysRevB.80.155302} {\bibfield  {journal} {\bibinfo
  {journal} {Phys. Rev. B}\ }\textbf {\bibinfo {volume} {80}},\ \bibinfo
  {pages} {155302} (\bibinfo {year} {2009})}\BibitemShut {NoStop}%
\bibitem [{\citenamefont {Duan}\ \emph {et~al.}(2018)\citenamefont {Duan},
  \citenamefont {Wang}, \citenamefont {Zheng}, \citenamefont {Wang},
  \citenamefont {Pan},\ and\ \citenamefont {Yang}}]{duan2018bulk}%
  \BibitemOpen
  \bibfield  {author} {\bibinfo {author} {\bibfnamefont {H.-J.}\ \bibnamefont
  {Duan}}, \bibinfo {author} {\bibfnamefont {C.}~\bibnamefont {Wang}}, \bibinfo
  {author} {\bibfnamefont {S.-H.}\ \bibnamefont {Zheng}}, \bibinfo {author}
  {\bibfnamefont {R.-Q.}\ \bibnamefont {Wang}}, \bibinfo {author}
  {\bibfnamefont {D.-R.}\ \bibnamefont {Pan}},\ and\ \bibinfo {author}
  {\bibfnamefont {M.}~\bibnamefont {Yang}},\ }\bibfield  {title} {\bibinfo
  {title} {Bulk rkky signatures of topological phase transition in silicene},\
  }\href {https://doi.org/10.1038/s41598-018-24567-w} {\bibfield  {journal}
  {\bibinfo  {journal} {Scientific reports}\ }\textbf {\bibinfo {volume} {8}},\
  \bibinfo {pages} {6185} (\bibinfo {year} {2018})}\BibitemShut {NoStop}%
\end{thebibliography}%
\end{document}